 \documentclass[final,1p,times]{elsarticle}

\usepackage{amssymb}

\usepackage[hyphens]{url}
\usepackage{hyperref}
\usepackage[anythingbreaks,hyphenbreaks]{breakurl}
\usepackage[nameinlink]{cleveref}
\usepackage{enumitem}
\usepackage{caption}
\usepackage{url}
\usepackage{array}
\usepackage{tabularx}
\usepackage{wrapfig}
\newcolumntype{P}[1]{>{\centering\arraybackslash}p{#1}}
\newcolumntype{Y}{>{\centering\arraybackslash}X} 

\journal{The Journal of Systems \& Software}

\begin{document}

\begin{frontmatter}


\title{Centralization potential of automotive E/E architectures}

\author[add1]{Lucas Mauser\corref{cor1}}

\cortext[cor1]{Corresponding author}
\ead{lucas.mauser@daimlertruck.com}

\author[add2,add3]{Stefan Wagner\corref{cor2}}

\affiliation[add1]{organization={Daimler Truck AG},
            city={Stuttgart},
            country={Germany}}
\affiliation[add2]{organization={Institute of Software Engineering, University of Stuttgart},
            country={Germany}}
\affiliation[add3]{organization={Technical University of Munich},
            city={Heilbronn},
            country={Germany}}

\begin{abstract}
Current automotive E/E architectures are subject to significant transformations: Computing-power-intensive advanced driver-assistance systems, bandwidth-hungry infotainment systems, the connection of the vehicle with the internet and the consequential need for cyber-security drives the centralization of E/E architectures. A centralized architecture is often seen as a key enabler to master those challenges. Available research focuses mostly on the different types of E/E architectures and contrasts their advantages and disadvantages. There is a research gap on guidelines for system designers and function developers to analyze the potential of their systems for centralization. The present paper aims to quantify centralization potential reviewing relevant literature and conducting qualitative interviews with industry practitioners. In literature, we identified seven key automotive system properties reaching limitations in current automotive architectures: busload, functional safety, computing power, feature dependencies, development and maintenance costs, error rate, modularity and flexibility. These properties serve as quantitative evaluation criteria to estimate whether centralization would enhance overall system performance. In the interviews, we have validated centralization and its fundament - the conceptual systems engineering - as capabilities to mitigate these limitations. By focusing on practical insights and lessons learned, this research provides system designers with actionable guidance to optimize their systems, addressing the outlined challenges while avoiding monolithic architecture. This paper bridges the gap between theoretical research and practical application, offering valuable takeaways for practitioners. 
\end{abstract}


\begin{highlights}
\item Practitioners rate centralization as a key enabler for software-defined vehicles
\item Centralization itself poses the risk of simply shifting a system's complexity
\item Increasing complexity and feature distribution negatively impact the error rate
\item Strict functional safety requirements can be met with local, embedded mini-ECUs
\item Legacy platforms limit development freedom and impede progress 
\end{highlights}


\end{frontmatter}


\section{Introduction}
\label{chap:Chapter1}
The distributed and embedded E/E architecture of the modern vehicle reaches its limits regarding multiple automotive system properties. Computing-power-intensive Advanced Driver-Assistance Systems (ADAS), bandwidth-hungry infotainment systems, the connectivity to services of the internet and the resulting need for cyber-security increase the complexity and drive a software-defined development approach \cite{bandur2021making}, \cite{RecentChallenges}. With regard to distributed and embedded E/E architectures, feature dependencies increase, scalability decreases, development and maintenance get more costly and less manageable as Vogelsang \cite{vogelsang2020feature} points out. The risk to lose control over the highly distributed system requires new approaches. Thus, automotive E/E-architectures have evolved from decentralized and distributed to domain-oriented architectures and are currently transforming into cross-domain architectures on their way to centralize within zonal E/E architectures, as \cite{bandur2021domain} and our previous work \cite{mauser2022methodical} show. The increasing functional logic and complexity of multiple distributed and embedded Electronic Control Units (ECUs) is being concentrated on a few, but powerful, High Performance Computers (HPCs). The embedded software design is evolving towards a centralized, service-oriented software design. Multi-core processing, mixed-criticality operating systems, virtualization, microservices, and containerization are examples of methods that enable this shift to a centralized, service-oriented HPC platform \cite{mauser2022methodical}.\vspace{2mm} 

Still, real-time and time-critical applications will coexist with the new service-oriented, event-driven systems according to Cinque et al. \cite{cinque2022certify}. First, there are simple applications that do not have complex and Central Processing Unit (CPU) intensive requirements. Second, there are applications that must meet strict time requirements to satisfy functional safety requirements according to ISO 26262. Thus, a system designer and function developer should be able to evaluate the benefit of a centralized implementation for each system and function before revising the overall E/E architecture of the vehicle in one iteration neglecting legacy applications. However, revising a system in one iteration is considered as challenging in the literature \cite{tappler2017model}. According to Conway’s Law, it can even require an organizational restructuring \cite{zerfowski2019functional}. Further research papers agree that the mindset of the employees, working strategies, team structures and culture must be adapted to evolve towards a software-defined company developing in line with DevOps practices \cite{zerfowski2019functional}, \cite{zerfowski2021building}.\vspace{2mm}

To resolve this conflict, many researchers focus on the different types of E/E architectures and contrast their advantages and disadvantages in their papers \cite{bandur2021making}, \cite{vogelsang2020feature}, \cite{bandur2021domain}, \cite{zerfowski2019functional}, \cite{kanajan2006exploring}. In addition, methods and tools that enable centralized and Service-Oriented Architectures (SOAs) are subject of past and present research \cite{zerfowski2019functional}, \cite{jesse2017future}, \cite{menard2020achieving}, \cite{bucaioni2020technical}, \cite{eisner2022systems}, \cite{RequEE}.\vspace{2mm} 

\begin{figure}[htbp]
	\centering		
	\noindent \includegraphics[width=0.8\linewidth]{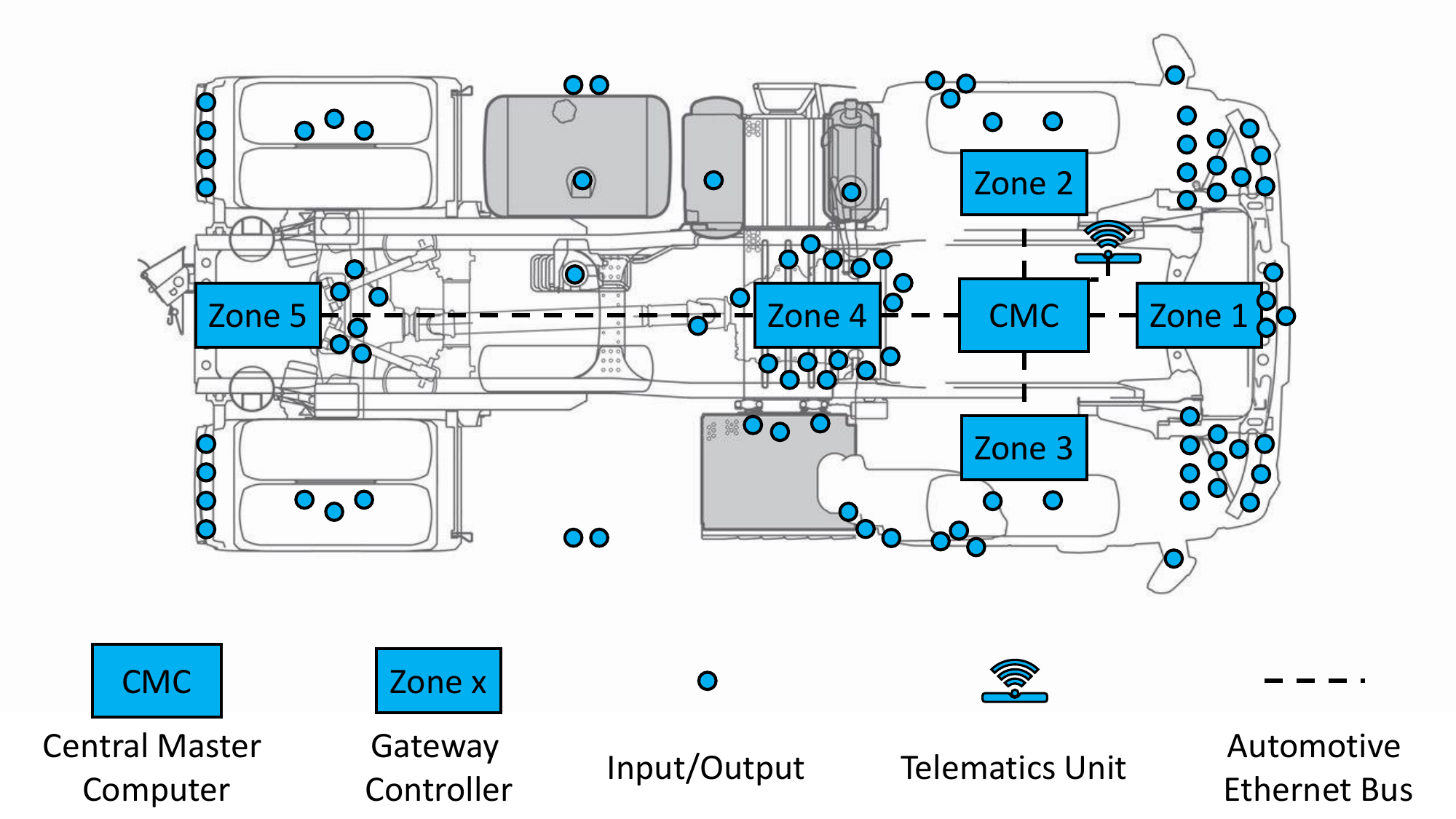}
	\caption{Zone-oriented E/E architecture approach}
	\label{fig:Zone-oriented}
\end{figure}

Parallels can be drawn with industrial production and manufacturing facilities, which have faced similar challenges in recent years. The massively increasing number of smart sensors and actuators developed for the Internet of Things (IoT) required and still requires efficient integration into a global ecosystem to improve productivity and provide added value for entrepreneurs and customers. To address this, the Industrial Internet of Things (I-IoT) and its 3-tier model has been developed to enable the “interconnection of anything, anywhere, and at any time” \cite{IIoT}, \cite{IIoT2}.
The I-IoT acts as a Cyber-Physical System (CPS), with the cyber systems providing control, networking, and computing, and the physical systems providing smart Inputs and Outputs (I/Os). 
The automotive, zone-oriented E/E architecture illustrated in \autoref{fig:Zone-oriented} follows a similar approach. Smart I/Os on physical layer receive and provide data as services on demand from and to a Central Master Computer (CMC) via, for example, publisher-subscriber concept using Automotive Ethernet Busses. The zonal gateway controllers gather the distributed I/Os to reduce the vehicle's overall cable harness length and by this the vehicle's overall weight \cite{CableLength}, \cite{CableLength2}, \cite{CableLength3}, \cite{SAEEE}. The gateway controllers serve as translators from and to the closely located smart I/O slaves for monitoring and control \cite{CableLength2}. The hardware-independent software logic is bundled on the application layer in the powerful CMC. The CMC is connected to a Telematics Unit, the importance of which is explained below.\vspace{3mm} 

\begin{figure}[htbp]
	\centering	
	\noindent \includegraphics[width=0.8\linewidth]{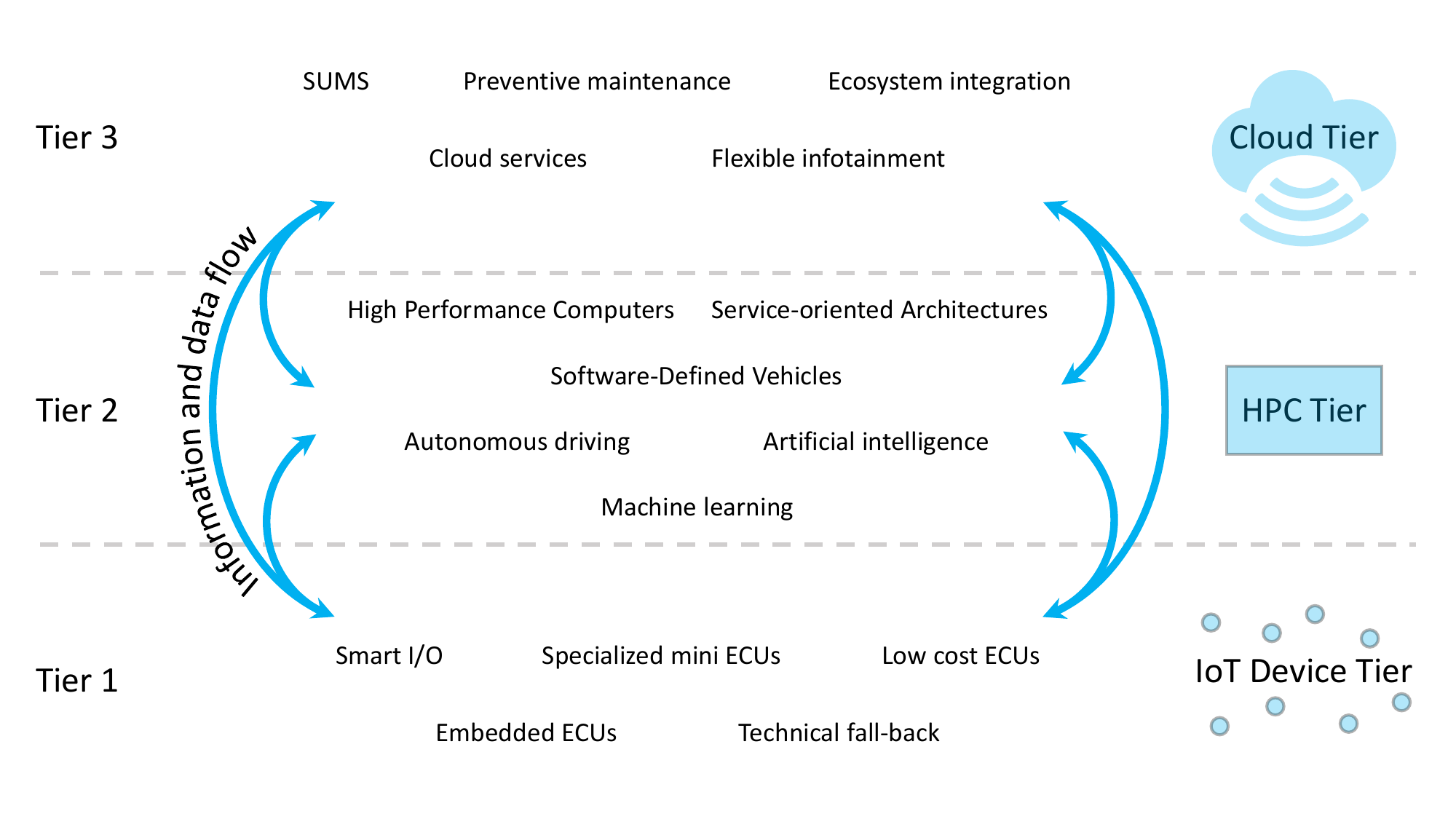}
	\caption{Classification of the zonal E/E architecture into the 3 tiers of I-IoT \cite{ConvergenceIT}}
	\label{fig:Zonal-IIoT}
\end{figure}

\begin{figure}[htbp]
	\centering		
	\noindent \includegraphics[width=0.75\linewidth]{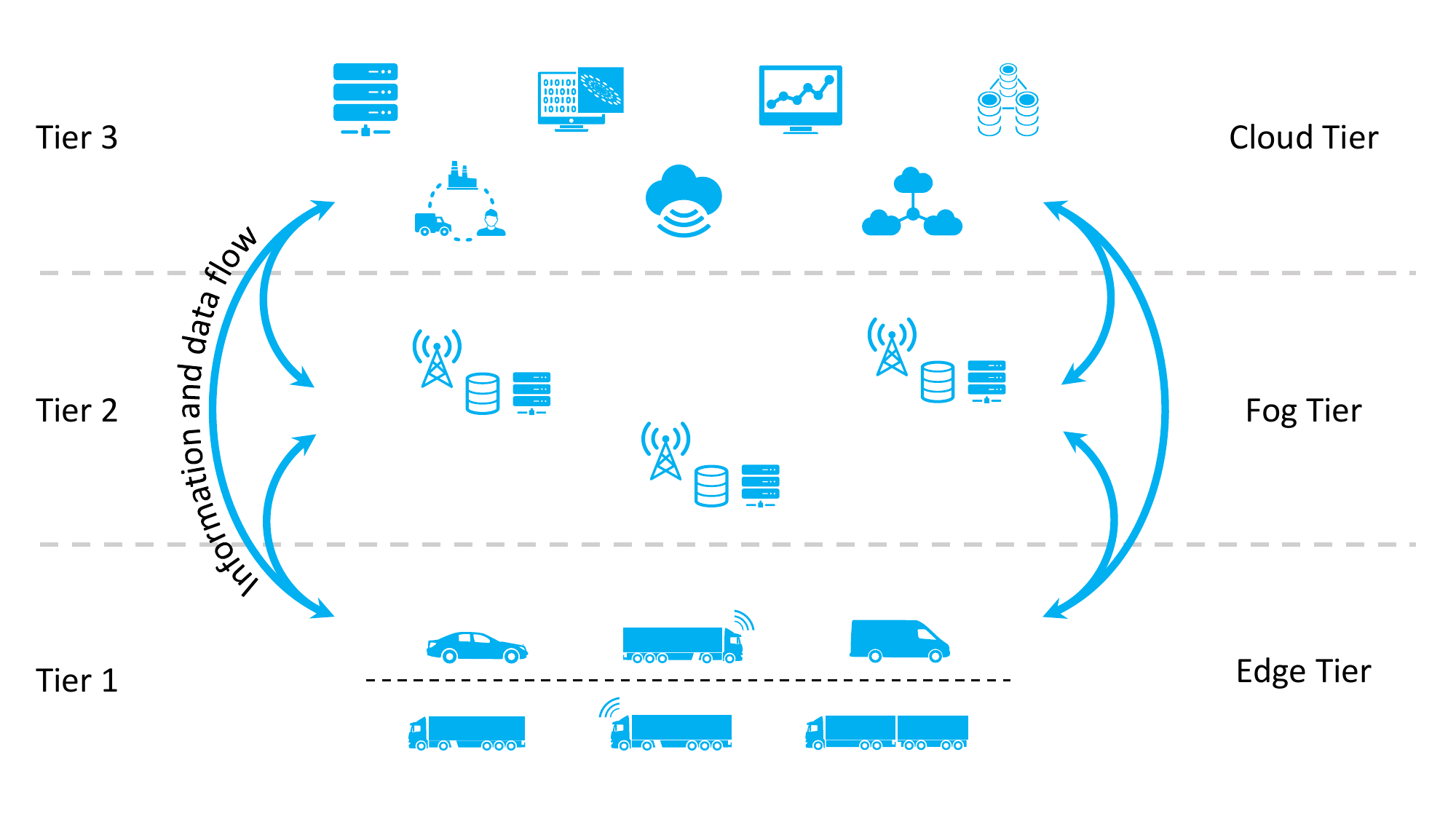}
	\caption{Classification of the vehicle as IoT device into the 3 tiers of I-IoT}
	\label{fig:IIoT}
\end{figure}

Here, the 3-tier model of I-IoT comes into play. It enables the creation of a transportation ecosystem, converging enterprise IT and product IT \cite{ConvergenceIT}. Looking at various reference architectures in the literature, the following three tiers best fit the automotive context: IoT device tier, edge tier, and cloud tier (see \autoref{fig:IIoT}) \cite{WeyrichEbert}, \cite{FraunhoferIIoT}, \cite{IIRA}, \cite{IOTA}, \cite{IEEEIoT}, \cite{TaxonomyIIoT}. Tier 1, the IoT device tier, represents the smart I/Os of the zonal E/E-architecture in \autoref{fig:Zone-oriented}. Tier 2, the edge tier, is also referred to as the HPC tier. This allows the reader to better understand the relationship between tier 2 and the CMC, or more generally, the HPC platform approach. Tier 3, the cloud tier, enables the connection of the vehicle to the internet and further instances. This connection is identified in the literature as the enabler of autonomous driving \cite{CloudEnablesAutonomous1}, \cite{CloudEnablesAutonomous2}. \autoref{fig:Zonal-IIoT} illustrates the classification and integration of the centralized, zonal E/E architecture of \autoref{fig:Zone-oriented} into the aforementioned 3 tiers to create added value for customers \cite{ConvergenceIT}.  

In particular, today's vehicle connectivity and emerging network technologies, such as 5G enable, for example, on-the-fly service deployment and Software Update Management Systems (SUMSs), or Vehicle-to-Vehicle communication (V2V). To create a transportation ecosystem that enables fully autonomous driving services, traffic management, road safety, data mining, and Vehicle-to-Everything (V2X), \cite{VEC} and \cite{CAV} introduce the additional Fog tier between the HPC and the cloud tier. It "enables low latency and reduced bandwidth consumption" compared to a computing within the cloud tier \cite{VEC}. More practically, it could be called the Vehicular Edge Computing (VEC) tier. \autoref{fig:IIoT} illustrates this approach.

Should or must the software logic be located on the I/Os themselves, on the edge tier, or centralized in a powerful cloud? I-IoT applications as well as automotive E/E architectures face the same decision to be taken \cite{IIoT}. As we will point out later, the specific use case and its requirements influence this decision for automotive E/E architectures.\vspace{2mm} 

Returning to automotive E/E architectures, a critical gap in current research is the absence of guidelines for system designers and function developers to assess the potential benefits of centralizing in-vehicle E/E architectures for enhancing overall system performance. To address this gap, it is essential to quantify this potential by linking it to specific evaluation criteria. The evaluation criteria developed within this paper represent typical automotive system properties currently constrained by emerging technologies as previously described. These criteria serve as benchmarks to analyze systems and function implementations, aiming to mitigate existing limitations and prepare for future technological advancements. 
This approach allowed us to formulate the first research question as a foundation for the interview study with the aim of supporting future system designers and function developers in their daily work:\vspace{2mm}  

\hypertarget{RQ1}{\textbf{RQ1:}} Which evaluation criteria can automotive engineers use to assess whether centralization of E/E architectures eliminates the automotive system limitations of today's distributed E/E architectures?\vspace{2mm}   


To gain insights into industry perspectives as our core contribution, we conducted a qualitative interview study on the identified evaluation criteria with practitioners at Daimler Truck AG. This study aimed to understand how practitioners perceive these evaluation criteria and their strategies in addressing the evolving technological challenges within the automotive sector. This investigation leads to our second research question:\vspace{2mm} 

\textbf{RQ2:} How do practitioners perceive and approach typical automotive system limitations?\vspace{2mm}  

The details and the setup of the interview study are described in \Cref{chap:Chapter3} in which we present the research method in detail. As the foundation for the interview questions, we identify and elaborate the evaluation criteria in the following section. 

\section{Background}
\label{chap:Chapter2}


In this section, we review relevant literature to identify key evaluation criteria for assessing the potential benefits of centralizing in-vehicle E/E architectures, which will complement our subsequent qualitative interviews with practitioners. In line with the practical focus of our research, this literature review was designed to provide preparatory groundwork to support our empirical interview study. We discuss the potential impact and threats to validity of our literature review in detail in \Cref{chap:ThreatsToValidity}, where we outline the steps taken to mitigate potential bias and limitations and the impact they may have on our findings.

\subsection{Busload}
\label{chap:Busload}
Systems require exchange of information towards other systems as also between its subfunctions and subcomponents. This fact brings us to the first evaluation criteria called busload. Distributed systems require communication to exchange information and increase the busload consequently as Kanajan et al. investigated in their research \cite{kanajan2006exploring}. In addition, the increasing complexity of modern vehicles also increases the busload of the communication channels. Cameras and LiDAR sensors with high resolution are examples for drivers of this trend \cite{RecentChallenges}, \cite{RequEE}, \cite{ADAS1}. Here, the time-driven Controller Area Network (CAN) protocol as the established communication protocol in automotive reaches its limit regarding its feasible payload \cite{bandur2021domain}, \cite{CANLimit}, \cite{TSN}. One approach to reduce high busloads is to switch from a signal-oriented communication towards a service-oriented communication like, for example, automotive Ethernet \cite{ServiceOriented}, \cite{AutosarAdaptive}. For automotive AUTOSAR, Sommer et al. \cite{sommer2013race} point out the Scalable service-Oriented MiddlewarE over IP (SOME/IP) with its publisher-subscriber concept. We see the risk that this approach addresses the symptom but not the cause with regard to a lean feature design that will be discussed in \Cref{chap:FeatDeps}. Thus, another approach to reduce high busloads is to centralize the system’s architecture and function distribution \cite{kanajan2006exploring}. It starts with the cause as it reduces or eliminates the need for communication between two ECUs.  

Building on \cite{kanajan2006exploring}, centralization is seen as a chance to reduce the busload and to eliminate system limitations. We define the busload as the first evaluation criterion. 

\subsection{Functional Safety}
\label{chap:FuSa}
Automotive systems are subject to functional safety according to ISO 26262. The norm defines different Automotive Safety Integrity Levels (ASIL) that depend on the consequences that a failure of a function will have. The risk is rated as a combination of the probability and the severity of the failure. It is used to define the ASIL. Based on the ASIL rating, Safety Goals (SGs) are defined. The system shall put itself into a safe state within the Fault Handling Time Interval (FHTI). The FHTI consists of the Fault Detection Time Interval for fault detection and qualification plus the Fault Reaction Time Interval (FRTI) to enter the safe state. The FHTI must be less than the Fault Tolerant Time Interval (FTTI), which is the span of time between the occurrence of a fault and the occurrence of the resulting hazardous event. \cite{iso26262} If a function is highly distributed, many components and subfunctions contribute to meeting the FHTI. The latencies increase as a result. By centralizing computing and I/O, this latency can be reduced \cite{kanajan2006exploring}. At the same time, it creates a single point of failure in the central HPC platform, which must be taken into account during the design phase, for example through hardware and software redundancy \cite{bandur2021domain}, \cite{SAEEE}. It is one reason why other publications consider distributed, domain-oriented E/E architectures to be more fault-tolerant \cite{CableLength}, \cite{FuSa1}, \cite{SAEEE}. Some even recommend dedicated, specialized ECUs for critical functions \cite{DedECUs}.    

In case a countermeasure against the single point of failure is feasible, centralization can optimize and reduce the system’s FHTIs. Thus, we see functional safety as a second evaluation criterion to benefit from system centralization. 

\subsection{Computing Power}
\label{chap:CompPow}
Driven by the increasing complexity in software and number of functions, the modern vehicle’s embedded ECUs require more and more computing power. Especially ADAS applications have to process high data amounts of different sensors and cameras \cite{ADAS1}, \cite{ADAS2}. As systems and functions get refined and extended over the development of one or more vehicle generations, the initial Microcontroller Unit (MCU) platforms are subject to the risk of reaching their computation limit. In particular, the software-defined emerging technologies mentioned previously drive this evolution. Optimization of task allocation, task prioritization, multi-core and many-core platforms are identified as short-term solutions in the literature \cite{claraz2014introducing}, \cite{macher2015automotive}, \cite{michel2016shared}. The introduction of a new MCU or HPC platform is identified as long-term solution. Such centralized, software-defined HPC platforms are seen as future-proof against today's technological challenges due to their high computing power and the resulting scalability. \cite{bandur2021making}, \cite{CloudEnablesAutonomous1} 

Thus, we rate centralization of the increasing amount of functions in HPC platforms as an approach to reduce the limitation by computing power of embedded ECUs. Consequently, we define computing power as the third evaluation criterion. 

\subsection{Feature Dependencies}
\label{chap:FeatDeps}
Features in automotive vehicles have become more and more connected to other features to provide innovative behavior \cite{broy2006challenges}. These dependencies can affect the design of the E/E architecture by considering the merging of domain controllers \cite{bandur2021domain}, \cite{mauser2022methodical}, \cite{CableLength2}. The number of intended and unintended dependencies is increasing constantly \cite{vogelsang2020feature}. According to Vogelsang’s related work investigation \cite{vogelsang2020feature}, feature dependencies complicate the system’s maintainability and the failure analysis. The risk of potential cross-effects and the many involved instances, either ECUs or Software Components (SWCs), reduce the manageability of the system and function developers. The chance to understand a perceived behavior and trace back failures to its root causes decreases. If a system evolved towards a monolith that is barely to maintain, a review and rework of the system, its features and the architecture can have a positive impact. Concepts as, for example, service orientation and the respective SOAs support to resolve an unmanageable monolith incrementally \cite{RethinkEE}. 

Feature dependencies within a system can be identified with different approaches as Vogelsang points out in \cite{vogelsang2020feature}. It mostly starts with a functional decomposition to break down functions to its basic elements. According to Vogelsang \cite{vogelsang2020feature} and Magnuson et al. \cite{RethinkEE}, functions and components used by many features have the potential to be integrated as a microservice with a common Application Programming Interface (API). This approach reduces complexity as a bigger task gets broken into many subtasks. The code has a better maintainability and issues can be isolated efficiently. Vogelsang \cite{vogelsang2020feature} recommends to bundle SWCs that are used by many features within a central “Platform Component Layer” (PLC). This refactoring approach shall help to reduce the number of components contributing to a feature dependency.  

Based on this, we identified centralization within a domain controller or an HPC platform as a countermeasure to reduce feature dependencies and thus the complexity of modern automotive systems. The system’s limitation by feature dependencies is defined as the fourth evaluation criterion.  

\subsection{Development and Maintenance Costs}
\label{chap:DevCosts}
As remarked in \Cref{chap:FeatDeps}, complexity and monolithic structures increase development and maintenance costs. The adding of new functions in an unmanageable monolithic environment is complicated as clear interfaces are missing and many feature dependencies need to be considered. The root causes of faulty behavior are difficult to trace back. There is also the risk that the system’s architecture did not receive a fundamental review and rework with continuous expansion. By this, the initially planned system does no longer satisfy the needs of the new requirements. In addition, backwards compatibility enforced by legacy functions affects the degree of freedom for such a rework of the system’s architecture \cite{RequEE}. As automotive projects have mostly a duration of several years, the stepwise implementation of new functions and new technologies is predetermined. A centralized approach can counteract those issues by ensuring portability and simplicity of function contributions and SWCs. As today's focus is often on cost optimization of embedded hardware, the overall software complexity increases, leading to excessive software development costs \cite{RethinkEE}. Here, the large-scale reuse of application software across "multi-product families and multi-brand environments" contributes to the reduction and amortization of development and maintenance costs \cite{SAEEE}, \cite{DedECUs}, \cite{RethinkEE}. Tools and methods like, for example, containerization or the usage of microservices ensure scalability and maintainability by reducing complexity \cite{kugele2018data}. 

As a result, we identified development and maintenance costs as another limitation in modern automotive architectures that are subject to continuous change and extension \cite{vogelsang2020feature}. The tools and methods of centralization support in reducing this limit. 

\subsection{Error Rate}
\label{chap:ErrorRate}
With increasing complexity and distribution of systems and its functions, error rate increases at the same time \cite{kanajan2006exploring}. The high number of feature dependencies between ECUs, functions and subfunctions increase the risk that changes of subfunctions impact others \cite{lotz2019microservice}, \cite{vogelsang2013feature}. Failures can cascade through the involved instances of a distributed system and even through a System of Systems (SoS) \cite{pelliccione2020beyond}.  

There are only a few articles on the relation between the distribution of an automotive system and its error rate as, for example, the one of Pelliccione et al. \cite{pelliccione2020beyond}. We build on the available literature and set up the preliminary thesis, that the centralization and service orientation of a system reduces its error rate. Thus, we define the error rate of a system as the sixth evaluation criterion to estimate the potential of centralization for the overall system’s performance. Nevertheless, new technologies and methods will result in new challenges and error patterns that need to be considered and mastered.    

\subsection{Modularity and Flexibility}
\label{chap:ModulAndFlex}
Modularity and flexibility of automotive systems gain in importance evolving towards a software-defined vehicle. Functions and applications shall be updateable over the air via telematics during the vehicle’s lifetime \cite{SAEEE}, \cite{scheer2020star3}. Literature rates modularity and flexibility of the typical E/E architecture differently. Kanajan et al. \cite{kanajan2006exploring} and Vignesh et al. \cite{SAEEE} analyzed a decentralized, distributed system as most flexible and modular as features and I/O can be handled by one dedicated component. Adding or removing features is feasible. This approach represents the component-based E/E architecture that literature rates as disadvantageous in modern vehicles due to multiple reasons \cite{bandur2021making}, \cite{sommer2013race}, \cite{lotz2019microservice}. Instead, modern vehicles shall profit of the advantages of feature-oriented E/E architectures to keep portability to new projects and be able to follow the emerging technologies. To do so, features must be defined technology-independent as described in beginning of \Cref{chap:Chapter2}. Combining this approach with a zone-oriented E/E architecture as illustrated exemplary in \autoref{fig:Zone-oriented}, system designers have an opportunity to split computing and I/O. This approach is similar to the decentralized I/O with centralized computing architecture of Kanajan et al. \cite{kanajan2006exploring} and Magnusson et al. \cite{RethinkEE}. Ensuring SOAs on SW level of the ECUs enable loose coupling of the SWCs. Modularity, flexibility, and portability is given so that the software can be centralized into the CMC \cite{mauser2022methodical}, \cite{jesse2017future}. The CMC can even offload tasks via the Telematics Unit to cloud servers so that, for example, computing-power-intensive software logic is run outside of the vehicle. While the interface to the I/O must be kept as generic as possible so that hardware of different technologies can be used \cite{RequEE}, \cite{lotz2019microservice}. The features in the form of application logic in the CMC can remain independent of the underlying technology of the I/O. By introducing containers in the CMC, functions can be added or removed modular and flexible as Kugele et al. point out \cite{kugele2018data}. By creating a zone-oriented architecture, special focus must be kept on the single point of failure within the CMC. Redundancy of the power supply, redundancy of the CMC itself or extensive software design and validation must be compliant with ISO 26262 and its corresponding ASIL rating \cite{iso26262}. 

We identify in literature a clear advantage of centralization with focus on the attributes' modularity and flexibility. As a result, we define modularity and flexibility as the seventh and last evaluation criterion to estimate the potential of centralization. \autoref{tab:Eval-Crit} gives a concluding overview of the within this section identified evaluation criteria. The criteria shall support system designers in analysis if their system is subject to limitations with regard to the upcoming emerging technologies. We suggest centralization as an approach to eliminate limitations in those fields. In the following, we describe our interview study that we apply to rate the evaluation criteria as also the centralization suggestion. 

\begin{table*}[htbp]
	\caption{Overview of Evaluation Criteria}
	\label{tab:Eval-Crit}
	\begin{tabularx}{\linewidth}{>{\centering\arraybackslash}p{0.1\linewidth}|>{\centering\arraybackslash}p{0.8\linewidth}}
		\hline
		\textbf{\#} & \textbf{Evaluation criterion}     \\ \hline
		1           & Busload                           \\ \hline
		2           & Functional safety                 \\ \hline
		3           & Computing power                   \\ \hline
		4           & Feature dependencies              \\ \hline
		5           & Development and maintenance costs \\ \hline
		6           & Error rate                        \\ \hline
		7           & Modularity and flexibility        \\ \hline
	\end{tabularx}
\end{table*}

\section{Research Method}
\label{chap:Chapter3}
Based on the theoretical background of \Cref{chap:Chapter2}, the quantifiable system properties of \autoref{tab:Eval-Crit} can serve as criteria to estimate centralization potential. In a next step, we identify whether centralization is associated as one potential approach to reduce the limitations in these system properties.  

\subsection{Interview Participants}
\label{chap:Participants}
During the sampling process of the potential participants for the interviews, first, we considered experts for systems affected by the recent trends ADAS, infotainment, connectivity, and proprietary operating systems. Second, we also considered experts for real-time systems and domains as, for example, the powertrain domain. Even those mainstream domains are subject to big changes due to their current electrification. We chose participants with an experience of at least 10 years or higher. Only participant 8 has less than 10 years of experience but is known for his expertise and has a Ph.D. in the field the participant is working on today. Furthermore, we chose participants that are seen as knowledge carriers within the company. An overview of the participants is illustrated in \autoref{tab:Inter-Part}. The table shows the participants’ roles, the domain in that the participants work and their years of experience. All the participants look back to multiple years of experience in automotive development at Original Equipment Manufacturers (OEMs) and Tier-1 suppliers. By this, the interview study takes place in a quite isolated field. We consider this fact in the discussion in \Cref{chap:ResultsAndDiscussion} and for potential follow-up studies in the outlook of \Cref{chap:ConclusionAndOutlook}. As a first step, we want to reflect the current situation of a well-established automotive manufacturer. The comparison with interview participants of other companies and industries will follow in a second step. A further comparison can also involve participants with fewer years of experience as they potentially bring in different mindsets, methods, or tools. 

\begin{table*}[htbp]
\caption{Overview of Interview Participants and Expertise}
\label{tab:Inter-Part}
	\begin{tabularx}{\linewidth}{>{\centering\arraybackslash}p{0.03\linewidth}|>{\centering\arraybackslash}p{0.316\linewidth}|Y|>{\centering\arraybackslash}p{0.22\linewidth}}
	\hline
	\textbf{ID} & \textbf{Participant role}	  & \textbf{Participant domain}  & \textbf{Years of experience}\\ \hline
	1           	& Function developer                  & Thermal management system	 & 10\\ \hline
	2           	& BSW architect                		  & Entire vehicle			 	 & 15\\ \hline
	3           	& Lead network communication          & Entire vehicle	 		 	 & 21\\ \hline
	4           	& Lead infotainment                   & Infotainment	 			 & 12.5\\ \hline
	5           	& Software architect DTOS 			  & Entire vehicle	 		 	 & 16\\ \hline
	6           	& Lead ADAS                           & Active safety	 			 & 30\\ \hline
	7           	& E/E architect        				  & Entire vehicle	 		 	 & 25\\ \hline
	8           	& Function developer        		  & Battery management system	 & 6.5\\ \hline
	           	& SW architect and function developer & Powertrain	 			 	 & 18\\ \hline
	10           	& System engineer        			  & Battery management system	 & 20\\ \hline
	\end{tabularx}
\end{table*}

We started the interview study with a set of more than the initially planned 10 participants, in total 12 potential participants. The purpose was to create a reserve and allow a review of the interview setup and the questions based on the first 2 participants. As we were facing no issues and uncertainties during the first interviews, we continued as started. While participant 11 has the same functional role as participant 10 and participant 12 works within the same domain as participant 3, we did no longer perform the interviews with participant 11 and 12. 

Before inviting the participants to the interviews as well as in the beginning of the interviews, we informed the participants that their interviews will be part of a qualitative research study with the aim of publishing it in a paper. We assured the participants the anonymization of the interviews. We gathered the verbal permission to publish a paper with the interviews’ content as a central component. In case one of the participants would have declined it in the beginning of the interview, we would not have integrated the related interview into the research study and thus also not published it. 

\subsection{Interview Questions}
\label{chap:Questions}
The interview questions aim at collecting countermeasures to dissolve limitations in the previously mentioned system properties. We integrate the expertise of long-term automotive developers to get an overview of potential approaches. This overview is used to validate and draw parallels to our centralization suggestion of \Cref{chap:Chapter2}. In addition, we analyze the awareness within the Daimler Truck AG for modern optimization approaches of the IT domain applied in automotive systems. Those were identified and worked out in our previous paper \cite{mauser2022methodical} as transferred approaches used for centralization in automotive architectures.  

The interview is structured in two parts. In part 1, the participants do not know about the aim of the research study and the centralization suggestion. The participants can present their approaches and solutions to face the identified system limitations without priming. Before part 2, the participants are introduced to the overall research topic. In part 2, specific questions related to centralization shall point out the awareness for it, its relevance and importance to face the recent transformation and future challenges. Due to the number of questions below, the duration of one interview was initially planned with 45 minutes. It turned out as sufficient. For nine of the ten interviews, the whole duration has been used for discussion.       

In part 1, each evaluation criterion is subject to a qualitative discussion with the below three-part structure: 

\begin{itemize}
 \item Limitation statement 
 \item Question 1: Which action(s) would you undertake in case…reaches its limit / increases / cannot be met…?
 \item Question 2: How would you prioritize your actions and for which reasons in case you identify more than one?
\end{itemize}

\autoref{tab:Limitation-Statements} shows the limitation statement and the Question 1 for each evaluation criterion of \autoref{tab:Eval-Crit}. Question 2 is neglected in \autoref{tab:Limitation-Statements} as it remains the same for each evaluation criterion. By this, part 1 of the interview is completed. We conclude that the questions are clearly based on and derived from the identified evaluation criteria of \Cref{chap:Chapter2}, whereas the definition of the questions in part 2 took place in a different way. We use part 1 to draw parallels between the participants’ approaches and our centralization suggestion for validation. Part 2 of the interview has a different purpose. As the participants are introduced into the research study’s topic before part 2, the questions are specifically focused on centralization and its consideration in automotive systems development. In doing so, we want to further investigate and validate centralization as an approach to make the present E/E architecture future-proof. In a first run, we want to understand if centralized approaches are already in use. This leads to the first question: 

\begin{itemize}
 \item Do you consider any kind of centralized approach during development of functions? 
\end{itemize}

We give special attention in the discussion about differences dependent on the participant’s domain. We expect that different environments and requirements demand different approaches. Centralization most probably is not the panacea for each use case due to, for example, strict FHTIs \cite{iso26262} or special requirements. Thus, we ask the participants about criteria to design a function considering its distribution in the overall system.  

\begin{itemize}
 \item Do you consider criteria to decide how centralized versus how distributed a function shall be structured or deployed?  
\end{itemize}

Still, literature rates the centralized architecture as a key enabler to master the modern technologies as pointed out in \Cref{chap:Chapter1}. We want to validate this thesis by the interview participants that are experienced automotive developers. Based on this, we defined the following question: 

\begin{itemize}
 \item Do you rate centralization as an important key factor to maintain a future-proof E/E architecture and for which reasons? 
\end{itemize}

To conclude the interview, we close the circle and come back again to the evaluation criteria of \Cref{chap:Chapter2}. The participants shall rate if centralization is a proper approach to eliminate system limitations in each of the evaluation criteria.  

\begin{itemize}
 \item  Do you rate centralization as supportive in resolving the limitations mentioned before? 
\end{itemize}

This question complements our analysis of part 1 of the interview and the parallels we draw to our centralization suggestion. The low number of participants does not fulfill a quantitative research study that can be part of a follow-up study. The detailed, qualitative investigation and comparison between the results of part 1 and part 2 will be part of the following section. 

\begin{table*}[htbp]
	\caption{Limitation Statement and Question 1 of Evaluation Criteria}
	\label{tab:Limitation-Statements}
	\begin{tabularx}{\linewidth}{>{\centering\arraybackslash}p{0.03\linewidth}|>{\centering\arraybackslash}p{0.4\linewidth}|Y}
\hline
\textbf{\#} & \textbf{Limitation statement}	  & \textbf{Question 1}  \\ \hline
1   & Busload reaches its limit                 					& Which action(s) would you undertake in case one of your communication channels reaches its busload limit?	 \\ \hline
2   & Required FHTI of a SG cannot be met                		    & Which action(s) would you undertake in case the FHTI of a SG cannot be met?			 	 \\ \hline
3   & Computing power reaches its limit         					& Which action(s) would you undertake in case the ECU you are responsible for reaches its maximum load?	 		 	 \\ \hline
4   & Feature dependencies and number of ECUs per feature increase  & Which action(s) would you undertake to reduce complexity?	 			 \\ \hline
5   & System development and maintenance costs increase over time   & Which action(s) would you undertake in case the development and maintenance of your system gets more and more costly?	 		 	 \\ \hline
6   & System becomes more and more error-prone over time            & Which action(s) would you undertake in case your system gets more and more error-prone?	 			 \\ \hline
7   & System becomes less modular and flexible over time        	  & Which action(s) would you undertake to keep modularity and flexibility in your system?	 		 	 \\ \hline
	\end{tabularx}
\end{table*}

\section{Results and Discussion}
\label{chap:ResultsAndDiscussion}
\subsection{Methodology of the Interview Analysis}
\label{chap:IntAnaly}

In the following, we will edit and prepare the results of the single interviews objectively to integrate them into our discussion, conclusion, and outlook. The basis for the analysis is the reporting of the answers. It supports to cluster similar responses and perform an inductive coding later. This allows us to derive patterns. Our aim is not to achieve statistical significance, but to provide an overview, to facilitate reading, and to draw qualitative conclusions from the patterns identified. We therefore accompany the reporting of answers with aggregating, tabular representations of the ranked, recurring patterns. We discuss individual occurrences if they were deduced in a plausible way. In this way, we satisfy the qualitative component of the present research. 
In addition, we present the interview questions from part 2 graphically to illustrate the reader tendencies. The arguments for the statements of the respective participant can be found in the following \Cref{chap:Results}, the reporting of the answers themselves.
       
\subsection{Results}
\label{chap:Results}
\textbf{Participant 1} is an experienced function developer in the thermal management system of the powertrain domain with an experience of more than ten years. The system itself is characterized by its high distribution among I/O as, for example, valves and compressors of the cooling circuits.  

To face the busload limit in the short term, P1 recommends to remove unneeded signals and reduce cycle times of certain frames. As a next step, P1 proposes the raising of the baud rate by, for example, updating a CAN 2.0 to a CAN-FD protocol. If the there is no more possibility to increase the baud rate in the used protocol variant, P1 faces the need to switch the physical layer or split the bus into multiple buses. The participant differs between short-term and long-term measures depending on time and costs that those consume. P1 recommends the split of the bus into multiple buses only for a next generation due to the need of additional pins. 

To meet the FHTI of a SG, P1 recommends the prioritization of dedicated tasks in the task handler. As a next step, P1 proposes to introduce multi-core and many-core. P1 emphasizes to keep an eye on the scheduling and allocation principle to reach a balanced load of the CPU cores.  

In case the computing power reaches its limit, P1 suggests to ensure an efficient coding style as complicated libraries in model-based development by Simulink lead to a bad runtime performance. In addition to that, P1 proposes overarching quality requirements for software developers. P1 also proposes, analogous to the FHTI, the integration of a multi-core or many-core processor. In the long term, P1 recommends to switch to a more powerful MCU. Also, for this limit, P1 differentiates between short-term and long-term actions. P1 emphasizes the consideration of costs and time. 

To reduce the complexity by a high number of feature dependencies and ECUs per feature, P1 recommends the reorganization and integration of a feature within one module. The more important is an optimal placement of the feature before implementation according to P1. To achieve this, P1 sees the need for a performant system and feature requirement process. A further proposal by P1 is to increase the scope of the domain controller. 

P1 faces increasing development and maintenance costs by creating software platforms and communality. As a result, software can be transferred to other products and markets easily without the need for a new hardware variant or ECU. 

In case a system becomes more and more error-prone over time, P1 proposes to introduce coding guidelines and structure functions in subfunctions. P1 mentions the principle “divide and conquer”. In addition, proper testing in form of unit tests, Software-in-the-Loop (SiL) tests, Hardware-in-the-Loop (HiL) tests and regression tests shall ensure the correctness of the system’s functions. In a last instance, P1 recommends the reengineering of software modules. P1 considers each of the actions as long-term. The participant prioritizes the coding guidelines.       

If the modularity and flexibility of a system decreased, P1 recommends to analyze if the market requirements changed. A further proposed approach is to overwork the system and its architecture by breaking down the functions into subfunctions. P1 emphasizes the importance to develop software in-house that gives the function developers more flexibility. P1 does not prioritize the actions but requests a restructuring of the design and development principles. 

As of now, P1 does not consider any kind of centralized approach when developing functional software. Criteria on how centralized versus how distributed a function shall be are already considered as the thermal management system relies on multiple local I/O that are connected to the domain controller by a kind of a gateway controller. P1 rates centralization as an important key factor for a future-proof E/E architecture. According to P1, it increases the company’s agility and it is estimated as basis for flexible release updates via Flashing Over The Air (FOTA) as also for the software’s portability to the next project. Anyhow, P1 emphasizes that centralization must happen in small steps due to the engineering and tool environment, big company structures and fragile systems. This conclusion of P1 is based on a current integration of two ECUs into one. 

\begin{table}[htbp]
	\caption{Busload reaches its limit - Identified patterns with at least 2 mentions}
	\label{tab:EvBusload-Patterns}
	\begin{tabularx}{\linewidth}{>{\raggedright\arraybackslash}p{0.7\linewidth}|>{\centering\arraybackslash}p{0.3\linewidth}}
		\hline
		\textbf{Pattern} & \textbf{Mentions}     \\ \hline
		Prioritization: time- and effort-oriented & 9   \\ \hline
		Bus clean-up and cycle time optimization & 8 \\ \hline
		Communication protocol and technology update & 8 \\ \hline
		E/E architecture update & 8 \\ \hline
		Busload monitoring and analysis & 4 \\ \hline
		Review of function distribution & 3 \\ \hline
		Legacy concerns	& 2 \\ \hline
	\end{tabularx}
\end{table}

\textbf{Participant 2} is an experienced Basic Software (BSW) expert developing the base layer and its requirements for the entire vehicle as a central instance. The BSW development follows predominantly the automotive AUTOSAR standards. 

To face the busload limit in the short term, P2 recommends an analysis of the communication that might result into a cycle time reduction. In the long term, P2 proposes to increase the baud rate by changing from CAN 2.0 to CAN-FD or reconnecting the ECUs as a last instance. If it does not help, P2 sees the need for a shift of functions into other ECUs. 

To meet the FHTI of a SG in the short term, P2 recommends to increase the cycle time of dedicated messages. In a next step, P2 proposes to analyze the different timing parts, see \Cref{chap:FuSa}, of the FHTI. A selection of proper failure indicators can help to reduce the FDTI. If still the FHTI cannot be met, P2 sees the need to review the required ECUs and centralize their functions within a central ECU. 

In case the computing power reaches its limit, P2 recommends to optimize and adjust the MCU settings like, for example, the task handling and prioritization, the interrupt time, the number of interrupts or the MCU memory interface. In a long term, multi-core MCUs or the next MCU generation is seen as an approach. In addition, the centralization of functions in an HPC platform is mentioned.  

To reduce the complexity caused by a high number of feature dependencies and ECUs per feature, P2 rates a system analysis as necessary. An overview gives an indication which and how many ECUs are involved in which functions. Based on the overview, P2 proposes to restructure the architecture and optimize the function distribution. In addition, standardization with regard to AUTOSAR is seen as essential to keep the software’s portability.  

Concerning increasing development and maintenance costs, P2 emphasizes the relation between complexity and maintenance costs. A review is recommended to analyze if the initial validation fits to the use cases. “Why are the issues now present? Is adding of functions increasing the complexity and thus the maintenance?” To reduce complexity, P2 sets the focus on standardization and a generic system. 

Faced with an increasingly error-prone system, P2 reacts by analysis and bug fixing in the short term. In the long term, P2 proposes to set the engineering focus on robustness and lifetime that must also be a part of the validation. P2 recommends to analyze the failure kind. P2 differentiates between hardware and software related failures. P2 concludes to design the software independent of the hardware.     

To keep the modularity and flexibility of a system, P2 emphasizes the need of software abstraction and design guidelines. APIs support to keep modularity by clearly defined interfaces according to P2. In addition, P2 proposes to review and rework the system and set the focus on a system-oriented development. 

P2 is involved in the development of the proprietary Daimler Truck Operating System (DTOS) for HPC platforms. As P2 does not develop functions, the participant does not consider criteria on how centralized versus how distributed a function shall be structured. P2 rates centralization as an important key factor for a future-proof E/E architecture and refers to the mentioned reasons.

\begin{table}[htbp]
\caption{FHTI of a SG cannot be met - Identified patterns with at least 2 mentions}
\label{tab:FHTI-Patterns}
\begin{tabularx}{\linewidth}{>{\raggedright\arraybackslash}p{0.7\linewidth}|>{\centering\arraybackslash}p{0.3\linewidth}}
\hline
\textbf{Pattern} & \textbf{Mentions}     \\ \hline
Prioritization: time- and effort-oriented & 7   \\ \hline
E2E signal path and timing analysis & 6 \\ \hline
Task prioritization and load optimization on CPU level & 5 \\ \hline
Review of system architecture and function distribution & 5 \\ \hline
Increasing of message cycle time & 3 \\ \hline
Review of SG partitioning & 3 \\ \hline
\end{tabularx}
\end{table}

\textbf{Participant 3} acts as the team lead for network communication with responsibility for the entire vehicle. 

In case the busload reaches its limit, P3 proposes to analyze why the initially planned busload target is exceeded. “Do different vehicle modes affect the busload or is measurement equipment involved?” To resolve the issue, P3 recommends decontenting in the form of removal of functions or ECUs and the partitioning of the E/E architecture by bus splits with respect to the ECUs that communicate the most. In the long term, P3 points out to check the feasibility of a technological update on the network technology. In addition, the bundling of functions within one ECU is identified as an approach as communication on busses also increases the latency. P3 emphasizes to consider the economic impact before each step.    

To meet the FHTI of a SG, P3 proposes in a first run to check if the FTTI was properly defined and if the FDTI or the FRTI can be decreased. As a next step, P3 sees the need to increase the cycle time and switch from a time-based communication towards an event-based communication. If the short-term actions are not sufficient, P3 recommends to review and rework the partitioning of the SG and a fusion of the functional distribution within one ECU.   

Based on the experience of P3, the root causes for computing power limitations are mostly runtime issues. Thus, P3 recommends the reduction of the runtime. P3 emphasizes the relation of runtime to RAM and ROM that requires an efficient tuning. In addition, the modification of the OS, the task mapping, the task prioritization and further are introduced as actions. Analog to other participants, a review and rework of the coding guidelines is proposed. In the long term, P3 sees the need to switch to the next MCU generation. 

To reduce complexity, P3 emphasizes the transparency about features and functions as also their distribution to avoid similar or duplicated functions. P3 recommends to use central HPC platforms and specialized ECUs at their required positions. P3 differs between two levels: software-defined HPC platforms on level one and hardware-driven ECUs on level 2. P3 concludes that transparency creates new concepts. 

In case development and maintenance costs increase, P3 sees the need to keep complexity low mentioning the "Keep It Simple, Stupid!" (KISS) principle, flexible functions, simple function contributions for simpler debugging and the avoidance of spaghetti code. A review and rework of the system is seen as necessary to identify the root causes of the complexity. P3 emphasizes that backwards compatibility reduces the potential for optimizations. HPC platforms are proposed to reduce complexity as functionality is centralized. P3 lists the passenger car approach to introduce new model lines with new architectures. It reduces the maintenance costs but at the same time increases the development costs. 

According to P3, failures are the result of complexity. Thus, P3 refers to the approaches in the question above. SWCs shall be designed with a low number of interfaces. P3 sees the need to standardize approaches within the company to avoid spontaneous patching. P3 emphasizes the need for systems engineering, the rework of the system’s architecture and its functions especially in case the number of features and functions increased.   

To keep modularity and flexibility of a system, P3 points out once more the necessity of systems engineering. In case the system gets too complex, the SoS approach is proposed to cut a system. On high level, P3 requests to define the Key Performance Indicators (KPIs) for modularity and flexibility to make those characteristics measurable. P3 emphasizes that typical approaches like, for example, the consideration of containerization and microservices or the integration of HPC platforms must be evaluated domain-specific to fulfill their specific requirements.  

P3 considers centralization during architectural change requests for new ECUs. In close collaboration with the requester, P3 identifies the need for a new ECU before its integration into the E/E architecture. P3 rates centralization as a key factor for a future-proof E/E architecture pointing out the characteristics portability, maintainability, extendibility, and simplicity.\vspace{2mm}    

\textbf{Participant 4} is the lead engineer for the digital Human-Machine Interface (HMI) in the infotainment domain and looks back on an experience of more than twelve years. 

In case a communication channels reaches its busload limit, P4 proposes the revision of the cycle times and a cleaning up in a short term. Mid-term, P4 sees the chance to use another communication channel. Over the long-term, P4 recommends to adapt the E/E architecture or perform a technology increment. According to P4, the prioritization depends on the criticality of the busload. 

To meet the FHTI of a SG, P4 proposes to optimize the CPU load. One recommended action is to optimize the task prioritization of ASIL tasks regarding CPU load and memory allocation. Another recommended action is to reduce the prioritization of other tasks. P4 prioritizes each of the actions same as all of them are rated as short-term actions. 

To face a computing power limit, P4 proposes to optimize processes and remove not required or low priority functions. In a next step, an MCU increment or an MCU change is recommended. In the long term, P4 identifies the integration of a new MCU platform as a possible action. 

In case the number of feature dependencies and ECUs per feature increase, P4 proposes to review and rework the architecture. By this, the logic of old and new functions can be integrated into new ECUs. P4 proposes to transform existing ECUs in hardware slaves.     

Regarding increasing development and maintenance costs, P4 points out with a negative rating the typical OEM approach to shift the development to low-cost countries. Instead, in-house software design increases the flexibility according to P4. Another important action for P4 is the design of software platforms that can be reused globally. 

To reduce the error rate of a system, P4 proposes to document and review issues. This is the basis to set up a lessons learned repository according to P4. In addition, P4 recommends external coaches for software quality to identify antipatterns and smells. In a mid-term perspective, P4 sees the need for in-house software architects that support the design process in early phases. As a last instance, P4 mentions the refactoring of the software as an approach.    

To keep a system modular and flexible, P4 proposes to modularize any ECU of the vehicle so that FOTA and further features are feasible in a universal way. To do so, P4 sees the need to rework the system and the introduction of a common operating system. 

P4 considers centralization during development. The HMI itself is developed as “a stupid slave”. It shall be a simple monitor and input interface to the driver. To consider how centralized versus how distributed a function shall be, P4 analyzes feature dependencies. Concluding, P4 rates centralization as an important key factor to maintain a future-proof E/E architecture due to the need for a modular design regarding updates. In addition to that, P4 mentions a need for faster updates and better maintainability.

\begin{table}[htbp]
\caption{Computing power reaches its limit - Identified patterns with at least 2 mentions}
\label{tab:CompPow-Patterns}
\begin{tabularx}{\linewidth}{>{\raggedright\arraybackslash}p{0.7\linewidth}|>{\centering\arraybackslash}p{0.3\linewidth}}
\hline
\textbf{Pattern} & \textbf{Mentions}     \\ \hline
Load optimization on MCU level & 7   \\ \hline
MCU HW update & 6 \\ \hline
Prioritization: time- and effort-oriented & 6 \\ \hline
Multi-core implementation & 4 \\ \hline
Introduction of coding or design guidelines & 4 \\ \hline
Removal or shifting of functions & 3 \\ \hline
Root cause analysis & 2 \\ \hline
Review of SW architecture & 2 \\ \hline
\end{tabularx}
\end{table} 

\textbf{Participant 5} works on the proprietary DTOS as a software architect. P5 looks back on an experience of more than 16 years. 

In case one of the communication channel reaches the busload limit, P5 proposes to remove unused or redundant signals in a short term. According to P5, it is also important to understand why the limit is reached. Normally, this should be “no surprise” as the busload is rated as something to be monitored constantly. In a next step, P5 proposes to increase the baud rate, switch the physical layer, or add central gateways. P5 emphasizes the risk of those actions as it can affect many ECUs. P5 also sees a benefit of a proper documentation and monitoring platform for the communication channels to avoid redundancy and further. Concluding, P5 points out the impact of specific requirements as the ones to realize Secure Onboard Communication (SecOC) on the payload. It is recommended to behave active instead of reactive. Nevertheless, P5 mentions the time criticality of the limit to prioritize properly. 

To meet the FHTI of a SG, P5 proposes to analyze the overall task timing within the ECU. Dedicated tasks can be executed on dedicated cores with high priority. To ensure processing in time, P5 proposes static task allocation. If those short-term actions are not sufficient, P5 proposes to review the partitioning on feature level and how distributed the signal paths are. 

In case the computing power reaches its limit, P5 proposes to review the functions running on the ECU. The priority of the different functions shall be rated. P5 recommends to define the operating modes and optimize them for full load. In addition, the use of design guidelines is introduced as beneficial for code efficiency. The participant differentiates between safety-related and comfort-related functions to define the priorities. 

To reduce the number of feature dependencies and number of ECUs per feature, P5 proposes to define clear feature interfaces as APIs. Complexity must be abstracted to different levels of services and subservices according to P5. 

To optimize development and maintenance costs, P5 sees the need to reduce variants. According to P5, the support of many variants increases the costs due to the product complexity. P5 proposes to standardize technologies and focus on higher volume approaches. Instead, P5 recommends partnerships with special industry players for dedicated use cases and smaller volumes. P5 concludes with the phrase “Know what you are good at and get help in what you are not.” and recommends to cooperate in fields in which one does not provide a lot of value.   

In case the system gets more and more error-prone, P5 wants to understand the source of the fault by analysis. “Is it an unforeseen use case?” P5 rates lessons learned approaches as beneficial. In general, P5 recommends to focus on understanding of the use cases and the system’s requirements. In addition, P5 emphasizes to include monitoring methods for fault detection. 

Concepts to maintain modularity and flexibility over time must be already integrated from the beginning according to P5. P5 points out to consider sufficient processing power resources from the very beginning. Furthermore, P5 mentions approaches like, for example, containerization, scalability in HPC platforms and I/O capabilities to keep a modular and flexible architecture. The participant prioritizes all actions equally from a long-term perspective. 

P5 considers centralization in its development by moving functionality as much as possible up into an HPC layer. The participant considers to separate computing and I/O. According to P5, the default for function development should be a centralized approach. P5 proposes decentralized approaches only for use cases with special needs from functional safety point of view and others. The participant rates centralization as an important key factor to be faster in development and deployment to update functions easily.

\begin{table}[htbp]
\caption{Feature dependencies and number of ECUs per feature increase - Identified patterns with at least 2 mentions}
\label{tab:FeatDep-Patterns}
\begin{tabularx}{\linewidth}{>{\raggedright\arraybackslash}p{0.7\linewidth}|>{\centering\arraybackslash}p{0.3\linewidth}}
\hline
\textbf{Pattern} & \textbf{Mentions}     \\ \hline
Review and optimization of feature distribution or E/E architecture & 6   \\ \hline
Centralization of feature distribution or E/E architecture & 6 \\ \hline
Standards to maintain portability (e.g. AUTOSAR) & 2 \\ \hline
Introduction of specialized, HW-driven mini-ECUs & 2 \\ \hline
\end{tabularx}
\end{table}

\textbf{Participant 6} is the lead engineer for the development of the active safety for Daimler Truck worldwide. P6 looks back on an experience of 30 years. Due to availability, P6 could answer the questions only partially. Some key notes are available and presented in the following.  

With a focus on ADAS, P6 assesses the busload limit as a global issue. A sufficiently powerful processor must also process the large amount of data on a communication channel in a timely manner. In addition, emerging technologies, such as 5G, are required to share the data with a further or redundant logic in the backend.   

To meet the FHTI of a SG, P6 rates a proper pre-design of the architecture and a proper system distribution as the basis. 

To not reach the computing power limit during a project, P6 rates a proper definition of the required payload in the beginning of the project as mandatory. In addition, P6 proposes multi-core processing, an optimized task distribution on the different cores and to keep a reserve for future tasks. 

In case of increased complexity, P6 points out the need to ensure and validate the functions. To achieve this, validation on the various types of HiLs is necessary. P6 identifies the increased complexity as the result of newly introduced protocols like SecOC, automotive Ethernet and further.  

P6 considers centralization in development since a longer time due to the need of high computing power for ADAS applications. The participant recommends to make AUTOSAR, the communication creation and the integration process more lean. In addition, P6 requests to unify the developing tool chain company-wide.

\begin{table}[htbp]
\caption{System development and maintenance costs increase over time - Identified patterns with at least 2 mentions}
\label{tab:DevMainCosts-Patterns}
\begin{tabularx}{\linewidth}{>{\raggedright\arraybackslash}p{0.7\linewidth}|>{\centering\arraybackslash}p{0.3\linewidth}}
\hline
\textbf{Pattern} & \textbf{Mentions}     \\ \hline
System and function review and rework & 4   \\ \hline
Root cause analysis of cost drivers & 3 \\ \hline
Standardization and generalization of system and technologies & 2 \\ \hline
Introduction of SW/HW platforms and communality & 2 \\ \hline
In-House development to increase flexibility	 & 2 \\ \hline
Introduction of a new E/E architecture & 2 \\ \hline
\end{tabularx}
\end{table}

\textbf{Participant 7} is the lead architect for the E/E architecture of the entire vehicle and looks back on an experience of more than 25 years.   

P7 proposes similar actions as the other participants to reduce the busload. The participant also suggests a review of the function distribution. At most, P7 wants to avoid the communication and consolidate it within one component. 

To meet the FHTI of a SG, P7 emphasizes to analyze and understand the end-to-end chain. Afterwards, cycle times can be increased or dedicated tasks can be prioritized. If those actions are not supportive, P7 sees the need to review and rework the function distribution.  

In case the computing power reaches its limit, P7 proposes to optimize the usage of the present controller from a software perspective. In a second step, P7 recommends a hardware upgrade within the MCU family. 

To manage complexity and feature dependencies, P7 emphasizes to not just shift or consolidate the complexity into a more powerful ECU as it is currently handled in automotive development. Complexity must be mastered and cannot be removed according to P7. Thus, P7 sees the need to analyze and understand the dependencies. In addition, P7 suggests to reduce variance and focus on the required functions.     

As well to face increasing development and maintenance costs, P7 emphasizes the understanding of the system, its interfaces and dependencies. In addition, P7 suggests to specify headroom for future use cases after Start of Production (SOP). 

In case the system gets more and more error-prone, P7 recommends to abstract and cut the system and its interfaces. According to P7, one root cause for the high error rate is a system that evolved bigger and bigger. His proposal is to divide and conquer. As a last instance, P7 recommends to review the system owner and its working methods. 

As well in case the system loses its modularity and flexibility, P7 proposes to review the system and increase the understanding for it. The participant identifies I/O extender, specialized ECUs and variants as actions to become modular and flexible again. P7 equates modularity with scalability while P7 equates flexibility with expandability as also the time required to react on a customer’s request.  

Participant 7 already considers centralization during development by reusing functions as kind of a service. To decide how distributed versus how centralized a function shall be designed, P7 emphasizes the demand for KPIs. Based on P7’s experience, it is difficult to follow a top-down design process consistently as the subsystems have dedicated requirements. P7 is the opinion that a centralized, consolidated E/E architecture is required for the future. Still, the central logic also must be manageable by, for example, modularization. Thus, P7 concludes that centralization itself will not be sufficient. A proper system design, distribution and understanding are further key factors.

\begin{table}[htbp]
\caption{System becomes more and more error-prone over time - Identified patterns with at least 2 mentions}
\label{tab:ErrProne-Patterns}
\begin{tabularx}{\linewidth}{>{\raggedright\arraybackslash}p{0.7\linewidth}|>{\centering\arraybackslash}p{0.3\linewidth}}
\hline
\textbf{Pattern} & \textbf{Mentions}     \\ \hline
Analysis and debugging & 5   \\ \hline
Consistent systems engineering & 4 \\ \hline
Testing and validation	& 3 \\ \hline
Refactoring of SW modules and functions & 2 \\ \hline
Introduction of coding guidelines & 2 \\ \hline
Introduction of microservice approaches & 2 \\ \hline
Consistent development process w/o patching & 2 \\ \hline
Design simple and flexible & 2 \\ \hline
Introduction of a lessons learned process & 2 \\ \hline
Prioritization: Coding guidelines & 2 \\ \hline
\end{tabularx}
\end{table}

\textbf{Participant 8} works as a component and function developer in the battery management system and looks back on an experience of more than six years.  

To reduce payload, P8 proposes similar actions to the participants above.  

To meet the FHTI of a SG, P8 proposes a review of the functional and technical safety concept. P8 mentions similar actions to the interviews before like task allocation and prioritization. From a hardware perspective, P8 proposes the use of high-performant sensors or actuators to reduce detection, qualification, or reaction time. 

In case the computing power reaches its limit, P8 identifies the shift of calculations to higher performant ECUs as a possible action. In general, the participant recommends to review and optimize the software architecture or to mount a higher performant MCU. 

P8 has not experienced feature dependencies as a limitation so far. In case the complexity is increasing, the participant proposes to evolve from a component-oriented E/E architecture to a domain-oriented E/E architecture. 

According to P8, faulty specifications and improper system designs paired with change requests over time are the root causes for high development and maintenance costs. Thus, a review of the system design and architecture is proposed.  

In case the system gets more error-prone, P8 proposes to simplify and modify related functions. In worst case, the faults are hardware related so that it requires hardware updates.  

Concerning modularity and flexibility of a system, P8 emphasizes the risk of function increments after SOP.  

P8 considers centralization already during development by moving software logic into the domain controller. Sensors and actuators are kept close to the battery if required. In what concerns safety-functions and sensor communication, P8 suggests mini-ECUs with restricted, dedicated functionality. To decide how centralized versus how distributed a function shall be developed, P8 analyzes the local dependencies. According to P8, centralization is one approach to keep an E/E architecture future-proof. The participant emphasizes the need for proper interfaces. P8 rates centralization as a cost saver using synergies and reducing overheads. 

\begin{table}[htbp]
\caption{System becomes less modular and flexible over time - Identified patterns with at least 2 mentions}
\label{tab:Modularity-Patterns}
\begin{tabularx}{\linewidth}{>{\raggedright\arraybackslash}p{0.7\linewidth}|>{\centering\arraybackslash}p{0.3\linewidth}}
\hline
\textbf{Pattern} & \textbf{Mentions}     \\ \hline
Review and rework of system and architecture & 5   \\ \hline
Consistent systems engineering incl. process & 3 \\ \hline
Introduction of microservice approaches & 3 \\ \hline
Design guidelines for adding of SWCs & 3 \\ \hline
Root cause analysis	& 2 \\ \hline
Introduction of containerization & 2 \\ \hline
Bundling of SWCs in HPC platforms & 2 \\ \hline
I/O capabilities in lower levels & 2 \\ \hline
\end{tabularx}
\end{table}

\textbf{Participant 9} is a software architect for the domain controller of the powertrain domain. P9 works on BSW functions in the lower layers of AUTOSAR as also on application level. The participant has an experience of 18 years in this field.  

Facing high busloads, P9 proposes similar actions to the participants above. The participant emphasizes backwards compatibility and prioritizes considering time and cost criticality. 

The questions about the case that an FHTI of a SG cannot be met were skipped. Participant 9 did not face that challenge yet in a direct way. 

If the computing power reaches its limit, P9 recommends to review the load distribution. “Which instance consumes how much? Is there any faulty implementation?” The participant highlights to optimize functions with multiple calls from different instances. If technically feasible, the frequency of an MCU can be increased as a next step according to P9. Afterwards, a multi-core implementation is considered by P9. In case the RAM reaches its limit, participant 9 suggests to shift the memory allocations from RAM into ROM. P9 emphasizes that an increased computing power also affects the memory size.  

In case feature dependencies and number of ECUs per feature increase, participant 9 suggests to reduce the number of ECUs by introducing an HPC platform in combination with clustering by port extenders. According to P9, the present development designs in a too embedded way with restricted KPIs.  

To decrease development and maintenance costs, P9 suggests “In-Housing” of the development. According to P9, the more instances and companies are involved in a development process, the more complicated and costly it gets. In addition, the participant emphasizes to stay attractive as an employer to reduce fluctuation of employees. 

In case the system gets more error-prone over time, participant 9 highlights the importance of guidelines for coding, modeling, signal naming and further. In design, the capability must be considered to keep an overview of the system and its functions. Meaningful testing, regression tests and feedback loops are an essential part of the development process according to P9. Those ensure that available functions keep their validity over time. P9 sets priority on the guidelines. 

To keep modularity and flexibility, P9 suggests to abstract functions. Especially, functions that are called multiple times shall be separated and refactored as microservices. The participant prioritizes to invest time in the beginning of the design phase for a proper software and function design. In addition, P9 recommends a strong cooperation between the specification and implementation instance of a project. 

During development, P9 considers the unification of functions. The participant creates libraries to design microservices for functions that are called from different instances. To define the distribution of a function, P9 analyzes if sensors and actuators require a dedicated location. Based on this, the participant evaluates the need for specialized ECUs. P9 does not rate centralization itself as sufficient to maintain a future-proof E/E architecture. According to P9, it depends on the specific functional contribution of a system that needs to be reviewed for the single functions. The participant illustrates its statement with former projects in which the inclusion of functions within one ECU resulted in a more error-prone behavior as no proper systems engineering was performed.

\begin{figure} [htbp]
	\centering	
	\noindent \includegraphics[width=0.8\linewidth]{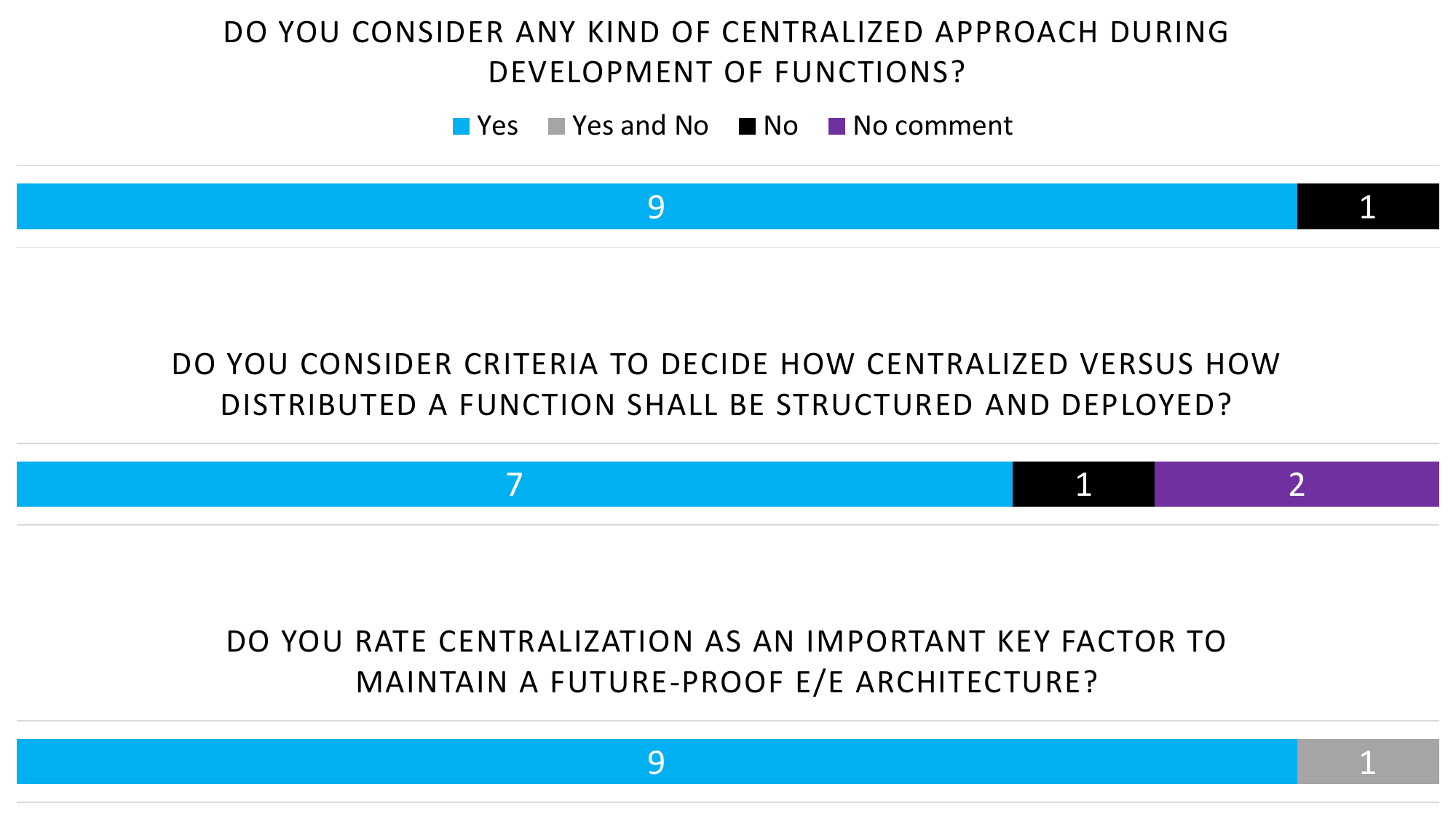}
	\caption{Results interview questions of part 2 - For arguments see textual reporting of answers in present \Cref{chap:Results}}
	\label{fig:Interview-Part2}
\end{figure}

\textbf{Participant 10} works as a system engineer for the battery management system. The participant gained in total 20 years of experience within this field.  

To reduce busload, the participant names similar actions to the other participants above. The participant prioritizes from short-term actions like a reduction of cycle times to long-term actions like an architecture update. P10 emphasizes legacy issues that have a high impact on the busload. The participant mentions a dump that is created over the time, as the engineers are afraid to remove no more required frames in case someone will need it one day. Thus, the big change comes with a new architecture according to P10. To avoid a busload overload during an architecture timeline, the participant suggests a central department tracking the busload. 

If the FHTI of a SG cannot be met, the participant recommends in a first run to review the FHTI itself if it is defined properly. If the SG is realized by a decomposition of measures, the participant proposes to review the decomposition. In general, P10 suggests a review of the system architecture including hardware and software both. This helps to find out at which position time can be reduced according to P10. Concluding, the participant emphasizes to analyze holistic and top-down on system level. Afterwards, the investigations can concretize. 

For P10, it is a known issue that the computing power of an ECU reaches its limit during its project timeline. The participant suggests the usage of logging modules in the BSW to trace the actual computing power and runtime. Furthermore, a debugger can be used to analyze timings and by this find out at which position most of the computing power gets lost. As soon as the computing hungry code and models are identified, the algorithms can be reviewed and optimized. At this point, P10 highlights similar to the other participants coding guidelines and static code analysis as actions. P10 differs between faulty implementations and not followed coding guidelines as root causes for high computing power losses. On a higher level, the participant suggests to review the software architecture and the requirements. According to P10, one must validate if the requirements are set up properly. P10 also names concrete actions like the switch-off of less important functions. 

Same as for the busload, P10 rates legacy issues as the root cause for increasing feature dependencies and number of ECUs per feature. The participant sees a chance to resolve this limitation by a new architecture. To do so, the present system architecture must be reviewed according to P10. It must be understood which functions belong to each other and which functions can potentially be combined in the same ECU. The participant emphasizes AUTOSAR as the basis to keep at least SWCs portable. P10 encourages using the chance to transfer functions. 

To reduce development and maintenance costs, participant 10 suggests to identify the cost drivers and review the system development process. A system rework is required according to P10 and in a last instance, the participant rates again a new architecture as a big chance. P10 proposes a vertical system development process to increase traceability. The participant names examples as the automotive SPICE standard. 

In case a system becomes more error-prone, P10 emphasizes the analysis of the failure kinds. The participant names single failures, systematic failures, and repeating failures as examples. Dependent on the kind of failures, measures must be derived. Based on his experience, P10 sees a high risk in patching as it decreases the system performance by creation of technical debts. As patching is a typical thing during development, it requires measures to overwork the technical debts as soon as they arise. 

Also, for a systems’ loss of modularity and flexibility, P10 sees legacy issues as one of the main root causes. According to P10, functional increments over the time that are not looped back on system level can decrease the modularity and flexibility of a system. Systems engineering and a related process is required to let the system grow properly in the opinion of P10. In addition, P10 highlights the usage of standards like AUTOSAR to keep modularity.  

As a system engineer, P10 evaluates in detail the function distribution within the battery management system. To consider how centralized versus how distributed a function shall be structured and deployed, P10 works out the ideal and merges it with the restrictions that are predefined. In general, participant 10 rates centralization as an important key factor to maintain a future-proof E/E architecture. According to P10, the high number of ECUs and interfaces increase the error rate. In addition, the costs increase due to redundant development, validation, and verification work. Nevertheless, the participant does not see centralization per se as the panacea for the discussed limitations. The high number of SWCs, their distribution, feature dependencies and complexity is only shifted into a central HPC platform ECU. They still need to be managed. Thus, P10 emphasizes systems engineering and a proper development process as the key. P10 highlights the challenge to merge the timelines of the many instances within a development process.


\begin{figure} [htbp]
	\centering	
	\noindent \includegraphics[width=0.8\linewidth]{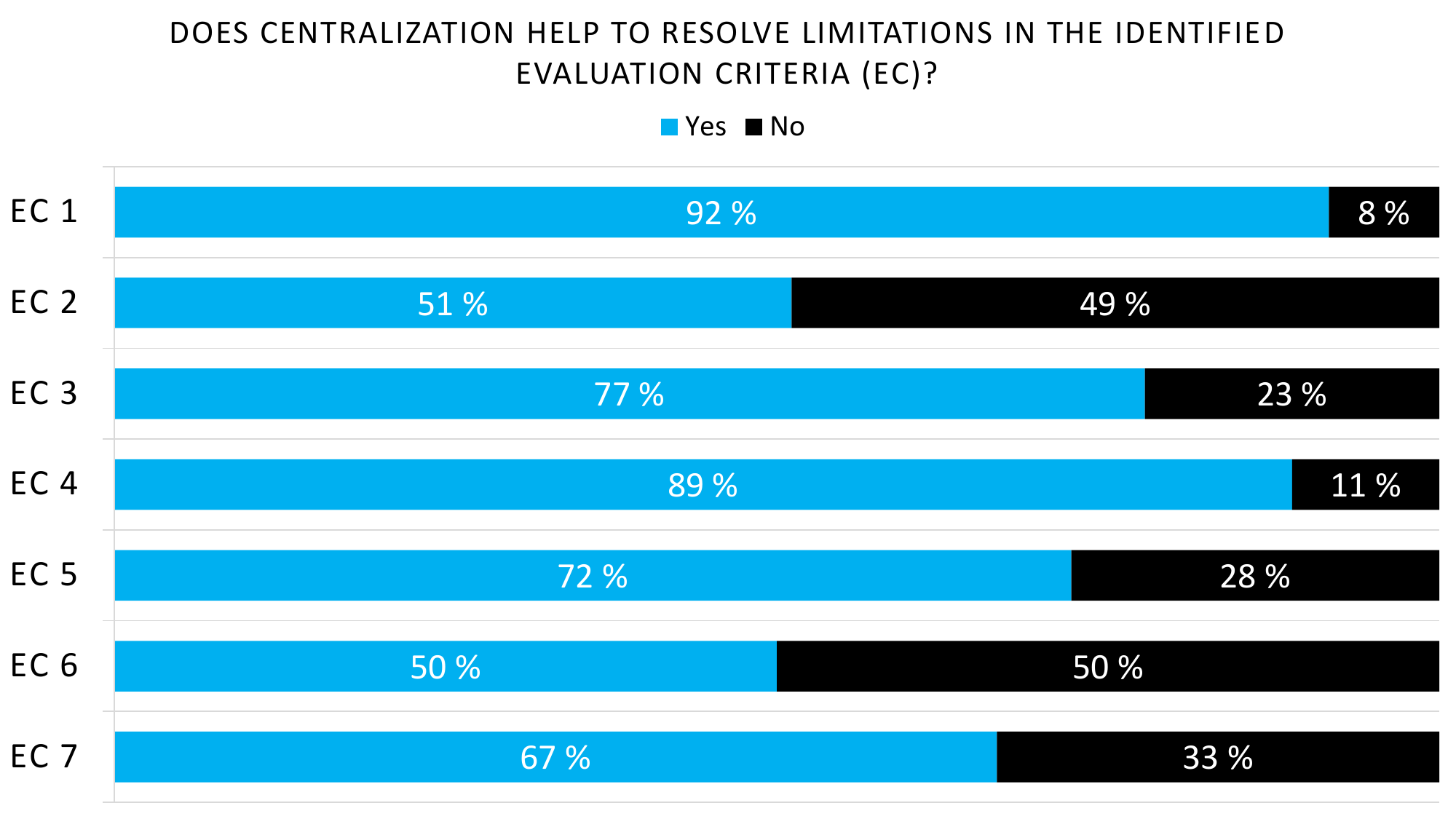}
	\caption{Results interview part 2 - Impact of centralization on evaluation criteria of \autoref{tab:Eval-Crit} displayed as means of answers rounded in percent according to following rating: Yes = 1, Partially = 0.75, Slightly = 0.6, Yes and No = 0.5, It depends = 0.5, To be analyzed = 0.5, Not necessarily = 0.5, No = 0}
	\label{fig:Interview-Part2}
\end{figure}

With each interview, we occasionally identify further interesting key notes and perspectives. However, the interview study achieved a balance of participants across departments, teams, and positions, so we continue with the discussion. To include more participants, we leave this as a subject for follow-up research and studies, as described in the outlook of this paper.

\subsection{Discussion}
\label{chap:Discussion}
In \Cref{chap:Chapter2}, we set up evaluation criteria that shall help in whether centralization can support to eliminate today’s automotive system limitations. As part of RQ1, the interview study of \Cref{chap:Chapter3} is used to validate the centralization approach as one potential solution to eliminate the limitations. In addition, the awareness for E/E architecture centralization and typical approaches of the engineers of an established OEM company are subject of RQ2. 

Facing the busload limitation, each of the participants tackles the challenge by a stepwise approach considering time-criticality and costs. Five of the ten participants also mention a centralized approach that considers the consolidation and shifting of functions in one ECU. This approach is mentioned in the last position as it is rated as a long-term action requiring much of preliminary work. Still, P7 emphasizes that the centralization will only shift the issue. It will also increase the communication within a potential HPC platform, in this case the inter-process communication (IPC). However, the IPC is rated much more far away from its load limit as current automotive bus systems are. None of the participants mentions explicitly the concept of service-oriented communication that we identified in literature as one approach to reduce high busloads. We interpret this fact as missing awareness among the participants for service-oriented communication protocols as automotive developers still focus on signal-oriented communication. 

As well to meet the FHTI of a SG, the by the participants suggested actions match. Some of the participants have more focus on MCU level and communication channel settings while other participants have more focus on the FTTI, FDTI and FRTI review as also a system and function distribution review. One participant recommends the integration of higher performant hardware to reduce the FDTI and FRTI. Another participant highlights a service-oriented approach to reduce latency by high busloads. Two participants mention a centralized approach as a potential action. Two further participants mention a more general review of the function distribution. Based on the previously mentioned conflict in literature (see \Cref{chap:FuSa}) and the interviews, we identified two future design approaches to satisfy safety requirements: 

\begin{itemize}
 \item Centralized approach that separates computing and I/O to reduce latency compared to a highly distributed end-to-end chain.
 \item Specialized mini-ECUs close to I/O with the shortest detection and reaction time for embedded application.
\end{itemize}
     
Depending on the required FTTI and the required computing power, the system designer can choose one of the designs. To face the computing power limit, the participants mention a high variety of actions. There are many different optimization approaches on MCU level. In a next step, the participants suggest the hardware upgrade within the MCU family or even to the next MCU generation. Those HW changes are rated as costly and long-term actions. In addition, some participants propose to introduce coding guidelines and quality requirements. Only P2 mentions centralization as a potential approach. P8 suggests the review and optimization of the software architecture at least. Contrary to literature, centralization is not always mentioned as an approach to reduce computing load limitations. Nevertheless, the clear majority of participants rated centralization as an approach to face computing power limitations after introduction in the research topic in part 2 of the interview. Thus, we rate centralization as an approach to face limited computing power as validated based on both the findings in the literature and the interviews themselves. 

Facing an increasing number of feature dependencies and components per feature, the majority of the participants proposes to abstract the system and to restructure the features inclusive their architecture with the goal to increase the scope of the domain controller. P2 highlights standardization to keep the portability of SWCs as a basis for the approach. P3 defines a level 1 represented by software-defined HPC platforms and a level 2 represented by hardware-driven ECUs as I/O. P4 also suggests to transform existing ECUs into simple hardware slaves similar to the zone-oriented approach of \cref{fig:Zone-oriented}. P7, P9 and P10 emphasize to not only shift or consolidate complexity in another ECU. Instead, dependencies must be understood and clearly defined to master the complexity. 

Only one participant mentions centralization as a potential approach to master increasing development and maintenance costs. Nevertheless, the majority of participants rated centralization as beneficial to face this development after introduction into the research topic in interview part 2. A further rating of the respective impacts of different centralization approaches on the development and maintenance costs is required. 

In case a system becomes more and more error-prone, the participants do not directly consider centralization as an approach. Anyhow, we identify single aspects of centralization as the participants propose to design software independent of the hardware and to apply the principle “divide and conquer”. Those aspects are highlighted in \Cref{chap:Chapter2}. Part 2 of the interview leaves split opinions on the centralization approach to face a system with a high error-rate. P7 emphasizes that even a centralized “software mountain” must be understood and properly designed. Based on the interviews, the basis for an efficient centralization is systems engineering. Systems must be abstracted into proper subsystems (SoS) as also functions into proper subfunctions. This approach is required for a distributed and for a centralized system equally to keep the error-rate low. Referring to the error rate research in \Cref{chap:ErrorRate}, the interviews support the thesis that complexity increases the error rate. This is an often-mentioned relation seen by the participants. 

According to the participants, modularity and flexibility requires conceptual work by systems engineering and standards like AUTOSAR as a basis. Many participants highlight approaches as containerization, microservices and the separation of computing and I/O to support the features and use cases of a modern vehicle. Part 2 of the interview validates this statement. 

In general, the participants approach the limitations analytically. They focus on the time-criticality that is an important aspect in ongoing projects. In the short term, the participants react. In the long term, the participants want to act by reviewing the system, its architecture, and its functions. Most of the participants are aware of centralization and associate it with advantages. Each of the participants rated centralization as an important key factor to maintain a future-proof E/E architecture. Still, the participants emphasize the importance of the conceptual work in the form of systems engineering.  

\subsection{Threats to Validity}
\label{chap:ThreatsToValidity}
While quantitative research focuses on generalizability, qualitative research focuses on particularizability \cite{erickson1985qualitative}. This implies a high variety of potential threats of validity owed by descriptive, interpretative, or explanatory contributions. Maxwell summarizes five threats to validity identified as commonly used in qualitative research \cite{maxwell1992understanding}. Due to its high reputation and dissemination in qualitative research, we want to use Maxwell's work to point out the threats to validity of our empirical study.

\subsection*{Descriptive Validity} 
The structure of our interview study helped us to ensure correct transcribing, see \Cref{chap:Questions}. The limitation statement was set up grammatically seen clear and comprehensible. Question 1 is of enumerable nature that avoids difficult interrelations per se. Question 2 let the participants repeat their actions and prioritize those so that we had a confirmation of question 1. After introducing the participants into the research study's topic, we asked in a clean and direct way about centralization in part 2 of the interview study. We gave attention to keep the questions grammatically simple with a clear focus on the participants' daily development work. During the interviews, we spoke with the participants in some unclear cases. After transcribing, we identified no abstruse or contradictory statements.

\subsection*{Interpretive Validity}
We attached importance to interview questions that do not influence the interviewee in its answers. Nevertheless, the interviews' results must be analyzed critically to exclude that the participants involved into future system architectures are blindly following the current trends. To do so, the questions of part 1 of the interview were quite generic and allowed a vast response spectrum. The interview atmosphere was trusting, and the arguments of the participants made it evident, that the participants did not gloss over or falsify their answers. 

\subsection*{Theoretical Validity}       
Based on current challenges in automotive development, we set up our research questions to finally derive supportive knowledge for daily practitioners. To ensure theoretical validity, also known as construct validity, appropriate research methods must be chosen so that the study measures what it claims to measure.

We started with a review of existing literature to identify current automotive system properties reaching their limitations and how those limitations are tackled. This review of existing literature with a practical scope described in \Cref{chap:Chapter2} did not follow a strict protocol for a Systematic Literature Review (SLR) which poses the risk of improper and missing relevant studies or evaluation criteria. Still, the identified evaluation criteria were aligned with the practitioners' experiences, as evidenced by the qualitative interview study. This alignment validates our findings to a certain degree despite it does not fully eliminate the theoretical risks associated with the non-rigorous literature review. To mitigate technical bias, we ensured the covering of both centralized and decentralized E/E architectures, avoiding favoritism towards either.         

In a second step, we set up a qualitative interview study with daily practitioners. The interview questions are based on the findings in existing literature and are constructed in a way that they clearly measure what we intend to measure (see \Cref{chap:Questions} and \Cref{chap:ThreatsToValidity}, descriptive and interpretive validity). 
Thus, the interview study as the core contribution builds on the review of existing literature and can be seen as an acknowledgement and practical extension of the findings in the literature. The attentive reader can compare her or his daily challenges with those of the present paper and pick out potential approaches to solve them.      
    
\subsection*{Generalizability} 
To ensure internal generalizability, we attached importance to a high variance of participants during the sampling process. The sampling covers experts from the domains impacted by the emerging technologies and current trends. In addition, we included experts from the mainstream domains that are also subject to big changes due to their electrification. The participants look back on an average working experience of 17.4 years reaching from 6.5 years to 30 years. Furthermore, we chose participants that are directly impacted by the system limitations and involved into discussions about advance development. 

The interview study included participants of only one single company, a long-established vehicle manufacturer. Nevertheless, we rate an intermediate external generalizability to other long-established OEMs including their Tier-1 suppliers. Excluding new, progressive players, automotive industry is still characterized by a similar domain-driven design that is currently on trial. Many of the companies are facing the same limitations worked out within this elaboration as they are facing the same emerging technologies \cite{bandur2021making}. The companies need to consider their legacy issues. In addition, we expect similar experience and expertise of the employees of other long-established OEMs and Tier-1 suppliers. Thus, we expect similar results of the same interview study. Instead, we expect differences in the applied systems engineering process within the different companies. We assume that the different companies tackle the challenges with different methods, processes, and tool chains. We exclude actively lower-level suppliers as those are characterized by specialist knowledge so that we cannot rate the external generalizability.

\subsection*{Evaluative Validity}  
Possibly, the evaluation criteria identified in the literature in \Cref{chap:Chapter2} do not cover all of the system properties that are reaching their limits due to the current automotive trends. Nevertheless, the set of seven properties reflects most of the present research papers and the interview participants did not mention further properties after introduction into the interview’s background within part 2 of the interview. Follow-up studies can easily extend their interview questions according to the questioning format of \autoref{tab:Limitation-Statements}.   

\section{Conclusion and Outlook}
\label{chap:ConclusionAndOutlook}
\subsection{Conclusion}
\label{chap:Conclusion}
In this paper, we have identified seven system limitations of automotive vehicles in their current technological transformation: busload, functional safety, computing power, feature dependencies, development and maintenance costs, error rate, modularity and flexibility. We worked out evaluation criteria to help system designers assess the benefits of centralized E/E architectures and to mitigate or even eliminate the above limitations. Validation was conducted through a qualitative interview study with ten experienced automotive development engineers and software architects. 

While the detailed opinions and responses of each interviewee can be found in the results section, we present the key takeaways below:

\begin{itemize}
\item \textbf{Centralization} - Most of the participants already consider centralized approaches during the development of their systems and functions. The participants rated centralization as an important key factor for a future-proof E/E architecture, emphasizing its improved updateability and flexibility. 
\item \textbf{Systems engineering} - The interview study highlighted the importance of systems engineering and high levels of abstraction to manage increasing complexity, following a System of Systems (SoS) approach. Without proper systems engineering, centralization risks shifting problems to the central instance or creating new challenges. 
\item \textbf{Functional safety} - We identified two architectural design approaches to meet varying strict functional safety requirements in E/E architectures. While a centralized approach can reduce latency compared to a distributed approach, specialized embedded mini-ECUs may still be required locally to achieve the shortest detection and reaction time.
\item \textbf{Error rate} - We were able to support the theory that increasing complexity and feature distribution can negatively impact the error rate of a system or function. Centralization, per se, was not identified as a countermeasure to reduce error rates but the need for systems engineering and functional decomposition. 
\item \textbf{Legacy issues} - Legacy issues must be considered during system rework, as they can limit development freedom and impede progress. Extensive changes can negatively impact reliability and functional safety of existing projects, favoring evolutionary rather than revolutionary changes to keep their strict timelines and mitigating risks. 
\item \textbf{Service-oriented communication} - We identified missing awareness for service-oriented communication by the participants while literature highlights it as an important approach to tackle busload limitations. 
\item \textbf{Development and maintenance costs} - Contrary to literature findings and assessments, participants were uncertain about the cost-saving potential of centralization in terms of development and maintenance costs.
\end{itemize}

\subsection{Outlook}
\label{chap:Outlook}
In future research that extends beyond the practice-focused scope of this study, we will apply a structured approach to deepen our literature review. Specifically, we plan to implement a systematic literature review (SLR) protocol to ensure comprehensiveness and transparency in identifying relevant studies. This will address the current methodological limitations and enhance the validity of our findings.

Additionally, we will expand our investigation to include the impact of different centralization approaches on development and maintenance costs. We also intend to extend our interview study to software and hardware engineers from the processor and IT industries. This will help us identify further potential approaches and methods for managing increasing complexity. There is a notable gap in automotive software architectures for high-performance computing (HPC) platforms. The consolidation of systems into microservices running on HPC platforms does not inherently reduce complexity. Therefore, we see the need for design guidelines to support systems and software architects during the transition from distributed to centralized architectures.

We believe there is significant potential to transfer approaches and concepts from the IT domain \cite{bogner2019microservices}, \cite{fritzsch2019monolith}, \cite{tyszberowicz2018identifying}. Including young graduates and research associates in the interview study may provide additional valuable insights, as they often approach challenges with fresh, independent perspectives.

Moreover, we aim to build on Vogelsang’s work \cite{vogelsang2020feature} to provide system designers with appropriate methods of functional decomposition to manage increasing complexity. As centralization progresses, designing measures to avoid single points of failure will become a critical aspect of our research.

\section*{Acknowledgements}
\label{chap:Acks}
We thank the participants of the interviews for their contribution in this research study, for the interesting input and discussions.
This research was partially supported by the German Federal Ministry of Education and Research in the project AutoDevSafeOps (01IS22087R) and the Baden-W\"urttemberg Ministry of Science, Research and the Arts in the Innovation Campus Future Mobility, projects SWUpCar and TESSOF.



	\bibliographystyle{elsarticle-num} 
	\bibliography{References}

\begin{thebibliography}{10}
\expandafter\ifx\csname url\endcsname\relax
  \def\url#1{\texttt{#1}}\fi
\expandafter\ifx\csname urlprefix\endcsname\relax\def\urlprefix{URL }\fi
\expandafter\ifx\csname href\endcsname\relax
  \def\href#1#2{#2} \def\path#1{#1}\fi

\bibitem{bandur2021making}
V.~Bandur, G.~Selim, V.~Pantelic, M.~Lawford, Making the case for centralized
  automotive e/e architectures, IEEE Transactions on Vehicular Technology
  70~(2) (2021) 1230--1245.
\newblock \href {https://doi.org/https://doi.org/10.1109/TVT.2021.3054934}
  {\path{doi:https://doi.org/10.1109/TVT.2021.3054934}}.

\bibitem{RecentChallenges}
G.~Xie, Y.~Li, Y.~Han, Y.~Xie, G.~Zeng, R.~Li, Recent advances and future
  trends for automotive functional safety design methodologies, IEEE
  Transactions on Industrial Informatics 16~(9) (2020) 5629--5642.
\newblock \href {https://doi.org/10.1109/TII.2020.2978889}
  {\path{doi:10.1109/TII.2020.2978889}}.

\bibitem{vogelsang2020feature}
A.~Vogelsang, Feature dependencies in automotive software systems: Extent,
  awareness, and refactoring, Journal of Systems and Software 160 (2020)
  110458.
\newblock \href {https://doi.org/https://doi.org/10.1016/j.jss.2019.110458}
  {\path{doi:https://doi.org/10.1016/j.jss.2019.110458}}.

\bibitem{bandur2021domain}
V.~Bandur, V.~Pantelic, M.~Dawson, A.~Schaap, B.~Wasacz, M.~Lawford, A
  domain-centralized automotive powertrain e/e architecture, Tech. rep., SAE
  Technical Paper (2021).
\newblock \href {https://doi.org/https://doi.org/10.4271/2021-01-0786}
  {\path{doi:https://doi.org/10.4271/2021-01-0786}}.

\bibitem{mauser2022methodical}
L.~Mauser, S.~Wagner, P.~Ziegler, Methodical approach for centralization
  evaluation of modern automotive e/e architectures, in: Software Architecture.
  ECSA 2022 Tracks and Workshops, Springer International Publishing, 2023, pp.
  165--179.
\newblock \href {https://doi.org/https://doi.org/10.1007/978-3-031-36889-9_13}
  {\path{doi:https://doi.org/10.1007/978-3-031-36889-9_13}}.

\bibitem{cinque2022certify}
M.~Cinque, L.~De~Simone, A.~Marchetta, Certify the uncertified: Towards
  assessment of virtualization for mixed-criticality in the automotive domain,
  in: 2022 52nd Annual IEEE/IFIP International Conference on Dependable Systems
  and Networks Workshops (DSN-W), IEEE, 2022, pp. 8--11.
\newblock \href {https://doi.org/https://doi.org/10.1109/DSN-W54100.2022.00012}
  {\path{doi:https://doi.org/10.1109/DSN-W54100.2022.00012}}.

\bibitem{tappler2017model}
M.~Tappler, B.~K. Aichernig, R.~Bloem, Model-based testing iot communication
  via active automata learning, in: 2017 IEEE International conference on
  software testing, verification and validation (ICST), IEEE, 2017, pp.
  276--287.
\newblock \href {https://doi.org/https://doi.org/10.1109/ICST.2017.32}
  {\path{doi:https://doi.org/10.1109/ICST.2017.32}}.

\bibitem{zerfowski2019functional}
D.~Zerfowski, A.~Lock, Functional architecture and e/e-architecture--a
  challenge for the automotive industry, in: 19. Internationales Stuttgarter
  Symposium: Automobil-und Motorentechnik, Springer, 2019, pp. 909--920.
\newblock \href {https://doi.org/https://doi.org/10.1007/978-3-658-25939-6_70}
  {\path{doi:https://doi.org/10.1007/978-3-658-25939-6_70}}.

\bibitem{zerfowski2021building}
D.~Zerfowski, S.~Antonov, C.~Hammel, Building the bridge between automotive sw
  engineering and devops approaches for automated driving sw development, in:
  Automatisiertes Fahren 2021: Vom assistierten zum autonomen Fahren 7.
  Internationale ATZ-Fachtagung, Springer, 2021, pp. 41--49.
\newblock \href {https://doi.org/https://doi.org/10.1007/978-3-658-34754-3_4}
  {\path{doi:https://doi.org/10.1007/978-3-658-34754-3_4}}.

\bibitem{kanajan2006exploring}
S.~Kanajan, C.~Pinello, H.~Zeng, A.~Sangiovanni-Vincentelli, Exploring
  trade-off's between centralized versus decentralized automotive architectures
  using a virtual integration environment, in: Proceedings of the Design
  Automation \& Test in Europe Conference, Vol.~1, IEEE, 2006, pp. 1--6.
\newblock \href {https://doi.org/https://doi.org/10.1109/DATE.2006.243895}
  {\path{doi:https://doi.org/10.1109/DATE.2006.243895}}.

\bibitem{jesse2017future}
B.~Jesse, M.~Weber, M.~Helmling, The future with soa, posix, tsn; automotive
  ethernet: Trends and challenges, Automobil-Elektronik 11 (2017) 32--35.

\bibitem{menard2020achieving}
C.~Menard, A.~Goens, M.~Lohstroh, J.~Castrillon, Achieving determinism in
  adaptive autosar, in: 2020 Design, Automation \& Test in Europe Conference \&
  Exhibition (DATE), IEEE, 2020, pp. 822--827.
\newblock \href
  {https://doi.org/https://doi.org/10.23919/DATE48585.2020.9116430}
  {\path{doi:https://doi.org/10.23919/DATE48585.2020.9116430}}.

\bibitem{bucaioni2020technical}
A.~Bucaioni, P.~Pelliccione, Technical architectures for automotive systems,
  in: 2020 IEEE International Conference on Software Architecture (ICSA), IEEE,
  2020, pp. 46--57.
\newblock \href {https://doi.org/https://doi.org/10.1109/ICSA47634.2020.00013}
  {\path{doi:https://doi.org/10.1109/ICSA47634.2020.00013}}.

\bibitem{eisner2022systems}
H.~Eisner, Systems engineering: Building successful systems, Springer Nature,
  2022.
\newblock \href {https://doi.org/https://doi.org/10.1007/978-3-031-79336-3}
  {\path{doi:https://doi.org/10.1007/978-3-031-79336-3}}.

\bibitem{RequEE}
H.~Zhu, W.~Zhou, Z.~Li, L.~Li, T.~Huang, Requirements-driven automotive
  electrical/electronic architecture: a survey and prospective trends, IEEE
  Access 9 (2021) 100096--100112.

\bibitem{IIoT}
H.~Xu, W.~Yu, D.~Griffith, N.~Golmie, A survey on industrial internet of
  things: A cyber-physical systems perspective, IEEE Access 6 (2018)
  78238--78259.
\newblock \href {https://doi.org/10.1109/ACCESS.2018.2884906}
  {\path{doi:10.1109/ACCESS.2018.2884906}}.

\bibitem{IIoT2}
J.~Lin, W.~Yu, N.~Zhang, X.~Yang, H.~Zhang, W.~Zhao, A survey on internet of
  things: Architecture, enabling technologies, security and privacy, and
  applications, IEEE Internet of Things Journal 4~(5) (2017) 1125--1142.
\newblock \href {https://doi.org/10.1109/JIOT.2017.2683200}
  {\path{doi:10.1109/JIOT.2017.2683200}}.

\bibitem{CableLength}
A.~Frigerio, B.~Vermeulen, K.~G.~W. Goossens, Automotive architecture
  topologies: Analysis for safety-critical autonomous vehicle applications,
  IEEE Access 9 (2021) 62837--62846.
\newblock \href {https://doi.org/10.1109/ACCESS.2021.3074813}
  {\path{doi:10.1109/ACCESS.2021.3074813}}.

\bibitem{CableLength2}
S.~P. Velusamy, M.~Y. Ghannam, H.~M. Kadry, Automotive sensor infrastructure -
  challenges and opportunities, in: 2022 IEEE International Symposium on
  Circuits and Systems (ISCAS), 2022, pp. 1018--1022.
\newblock \href {https://doi.org/10.1109/ISCAS48785.2022.9937749}
  {\path{doi:10.1109/ISCAS48785.2022.9937749}}.

\bibitem{CableLength3}
H.~M. Kadry, A.~Gupta, J.~M. Lawlis, M.~Volpone, Electrical architecture and
  in-vehicle networking: Challenges and future trends, in: 2022 IEEE
  International Symposium on Circuits and Systems (ISCAS), 2022, pp.
  1009--1013.
\newblock \href {https://doi.org/10.1109/ISCAS48785.2022.9937481}
  {\path{doi:10.1109/ISCAS48785.2022.9937481}}.

\bibitem{SAEEE}
N.~Vignesh, M.~Kumar, B.~Achuthan, S.~Badade, P.~Shivakumar, A.~Keshri,
  Centralized e/e architecture and evolution, Tech. rep., SAE Technical Paper
  (2023).

\bibitem{ConvergenceIT}
C.~Ebert,
  \href{https://consulting.vector.com/int/en/download/from-silos-to-synergy-industry-practice-for-converging-systems/}{From
  silos to synergy: Industry practice for converging systems}, Tech. rep.,
  accessed: 2024-03-30.
\newline\urlprefix\url{https://consulting.vector.com/int/en/download/from-silos-to-synergy-industry-practice-for-converging-systems/}

\bibitem{WeyrichEbert}
M.~Weyrich, C.~Ebert, Reference architectures for the internet of things, IEEE
  software 33~(1) (2015) 112--116.
\newblock \href {https://doi.org/10.1109/MS.2016.20}
  {\path{doi:10.1109/MS.2016.20}}.

\bibitem{FraunhoferIIoT}
M.~Heidrich, J.~J. Luo, Industrial internet of things (2016).
\newblock \href {https://doi.org/10.24406/publica-fhg-297578}
  {\path{doi:10.24406/publica-fhg-297578}}.

\bibitem{IIRA}
I.~I. Consortium, et~al., The industrial internet reference architecture,
  accessed: 2024-03-24 (2022).

\bibitem{IOTA}
J.~De~Loof, C.~M. SAP, S.~Meissner, A.~Nettstr{\"a}ter, A.~O. CEA, M.~T. SAP,
  J.~W. Walewski, Internet of things-architecture iot-a deliverable d1. 5-final
  architectural reference model for the iot v3. 0, IoT-A (257521) (2013)
  1--499.

\bibitem{IEEEIoT}
Ieee standard for an architectural framework for the internet of things (iot),
  IEEE Std 2413-2019 (2020) 1--269\href
  {https://doi.org/10.1109/IEEESTD.2020.9032420}
  {\path{doi:10.1109/IEEESTD.2020.9032420}}.

\bibitem{TaxonomyIIoT}
P.~Jayalaxmi, R.~Saha, G.~Kumar, N.~Kumar, T.-H. Kim, A taxonomy of security
  issues in industrial internet-of-things: Scoping review for existing
  solutions, future implications, and research challenges, IEEE Access 9 (2021)
  25344--25359.
\newblock \href {https://doi.org/10.1109/ACCESS.2021.3057766}
  {\path{doi:10.1109/ACCESS.2021.3057766}}.

\bibitem{CloudEnablesAutonomous1}
B.~Deng, J.~Nan, W.~Cao, W.~Wang, A survey on integration of network
  communication into vehicle real-time motion control, IEEE Communications
  Surveys \& Tutorials 25~(4) (2023) 2755--2790.
\newblock \href {https://doi.org/10.1109/COMST.2023.3295384}
  {\path{doi:10.1109/COMST.2023.3295384}}.

\bibitem{CloudEnablesAutonomous2}
N.~S. Tany, S.~Suresh, D.~N. Sinha, C.~Shinde, C.~Stolojescu-Crisan,
  R.~Khondoker, Cybersecurity comparison of brain-based automotive electrical
  and electronic architectures, Information 13~(11) (2022) 518.

\bibitem{VEC}
L.~Liu, C.~Chen, Q.~Pei, S.~Maharjan, Y.~Zhang, Vehicular edge computing and
  networking: A survey, Mobile networks and applications 26 (2021) 1145--1168.
\newblock \href {https://doi.org/10.1007/s11036-020-01624-1}
  {\path{doi:10.1007/s11036-020-01624-1}}.

\bibitem{CAV}
S.~Gupta, C.~Maple, R.~Passerone, An investigation of cyber-attacks and
  security mechanisms for connected and autonomous vehicles, IEEE Access
  (2023).
\newblock \href {https://doi.org/10.1109/ACCESS.2023.3307473}
  {\path{doi:10.1109/ACCESS.2023.3307473}}.

\bibitem{kitchenham2007guidelines}
B.~Kitchenham, S.~Charters, et~al., Guidelines for performing systematic
  literature reviews in software engineering (2007).

\bibitem{ADAS1}
H.~Askaripoor, M.~Hashemi~Farzaneh, A.~Knoll, E/e architecture synthesis:
  Challenges and technologies, Electronics 11~(4) (2022) 518.

\bibitem{CANLimit}
S.~Chakraborty, M.~Lukasiewycz, C.~Buckl, S.~Fahmy, N.~Chang, S.~Park, Y.~Kim,
  P.~Leteinturier, H.~Adlkofer, Embedded systems and software challenges in
  electric vehicles, in: 2012 Design, Automation \& Test in Europe Conference
  \& Exhibition (DATE), IEEE, 2012, pp. 424--429.

\bibitem{TSN}
M.~Ashjaei, L.~L. Bello, M.~Daneshtalab, G.~Patti, S.~Saponara, S.~Mubeen,
  Time-sensitive networking in automotive embedded systems: State of the art
  and research opportunities, Journal of systems architecture 117 (2021)
  102137.

\bibitem{ServiceOriented}
M.~Rumez, D.~Grimm, R.~Kriesten, E.~Sax, An overview of automotive
  service-oriented architectures and implications for security countermeasures,
  IEEE access 8 (2020) 221852--221870.

\bibitem{AutosarAdaptive}
S.~F{\"u}rst, M.~Bechter, Autosar for connected and autonomous vehicles: The
  autosar adaptive platform, in: 2016 46th annual IEEE/IFIP international
  conference on Dependable Systems and Networks Workshop (DSN-W), IEEE, 2016,
  pp. 215--217.

\bibitem{sommer2013race}
S.~Sommer, A.~Camek, K.~Becker, C.~Buckl, A.~Zirkler, L.~Fiege, M.~Armbruster,
  G.~Spiegelberg, A.~Knoll, Race: A centralized platform computer based
  architecture for automotive applications, in: 2013 IEEE International
  Electric Vehicle Conference (IEVC), IEEE, 2013, pp. 1--6.
\newblock \href {https://doi.org/https://doi.org/10.1109/IEVC.2013.6681152}
  {\path{doi:https://doi.org/10.1109/IEVC.2013.6681152}}.

\bibitem{iso26262}
Iso 26262: Road vehicles - functional safety, Tech. rep. (2018).

\bibitem{FuSa1}
K.~Jo, J.~Kim, D.~Kim, C.~Jang, M.~Sunwoo, Development of autonomous car—part
  i: Distributed system architecture and development process, IEEE Transactions
  on Industrial Electronics 61~(12) (2014) 7131--7140.

\bibitem{DedECUs}
M.~Broy, I.~H. Kruger, A.~Pretschner, C.~Salzmann, Engineering automotive
  software, Proceedings of the IEEE 95~(2) (2007) 356--373.

\bibitem{ADAS2}
S.~P. Velusamy, M.~Y. Ghannam, H.~M. Kadry, Automotive sensor
  infrastructure-challenges and opportunities, in: 2022 IEEE International
  Symposium on Circuits and Systems (ISCAS), IEEE, 2022, pp. 1018--1022.

\bibitem{claraz2014introducing}
D.~Claraz, F.~Grimal, T.~Leydier, R.~Mader, G.~Wirrer, Introducing multi-core
  at automotive engine systems, in: Embedded Real Time Software and Systems
  (ERTS2014), 2014.

\bibitem{macher2015automotive}
G.~Macher, A.~H{\"o}ller, E.~Armengaud, C.~Kreiner, Automotive embedded
  software: Migration challenges to multi-core computing platforms, in: 2015
  IEEE 13th International Conference on Industrial Informatics (INDIN), IEEE,
  2015, pp. 1386--1393.
\newblock \href {https://doi.org/https://doi.org/10.1109/INDIN.2015.7281937}
  {\path{doi:https://doi.org/10.1109/INDIN.2015.7281937}}.

\bibitem{michel2016shared}
L.~Michel, T.~Flaemig, D.~Claraz, R.~Mader, Shared sw development in multi-core
  automotive context, in: 8th European Congress on Embedded Real Time Software
  and Systems (ERTS 2016), 2016.

\bibitem{broy2006challenges}
M.~Broy, Challenges in automotive software engineering, in: Proceedings of the
  28th international conference on Software engineering, 2006, pp. 33--42.
\newblock \href {https://doi.org/https://doi.org/10.1145/1134285.1134292}
  {\path{doi:https://doi.org/10.1145/1134285.1134292}}.

\bibitem{RethinkEE}
A.~Magnusson, L.~Laine, J.~Lindberg, Rethink ee architecture in automotive to
  facilitate automation, connectivity, and electro mobility, in: Proceedings of
  the 40th International Conference on Software Engineering: Software
  Engineering in Practice, 2018, pp. 65--74.

\bibitem{kugele2018data}
S.~Kugele, D.~Hettler, J.~Peter, Data-centric communication and
  containerization for future automotive software architectures, in: 2018 IEEE
  International Conference on Software Architecture (ICSA), IEEE, 2018, pp.
  65--6509.
\newblock \href {https://doi.org/https://doi.org/10.1109/ICSA.2018.00016}
  {\path{doi:https://doi.org/10.1109/ICSA.2018.00016}}.

\bibitem{lotz2019microservice}
J.~Lotz, A.~Vogelsang, O.~Benderius, C.~Berger, Microservice architectures for
  advanced driver assistance systems: A case-study, in: 2019 IEEE International
  Conference on Software Architecture Companion (ICSA-C), IEEE, 2019, pp.
  45--52.
\newblock \href {https://doi.org/https://doi.org/10.1109/ICSA-C.2019.00016}
  {\path{doi:https://doi.org/10.1109/ICSA-C.2019.00016}}.

\bibitem{vogelsang2013feature}
A.~Vogelsang, S.~Fuhrmann, Why feature dependencies challenge the requirements
  engineering of automotive systems: An empirical study, in: 2013 21st IEEE
  International Requirements Engineering Conference (RE), IEEE, 2013, pp.
  267--272.
\newblock \href {https://doi.org/https://doi.org/10.1109/RE.2013.6636728}
  {\path{doi:https://doi.org/10.1109/RE.2013.6636728}}.

\bibitem{pelliccione2020beyond}
P.~Pelliccione, E.~Knauss, S.~M. {\AA}gren, R.~Heldal, C.~Bergenhem, A.~Vinel,
  O.~Brunneg{\aa}rd, Beyond connected cars: A systems of systems perspective,
  Science of Computer Programming 191 (2020) 102414.
\newblock \href {https://doi.org/https://doi.org/10.1016/j.scico.2020.102414}
  {\path{doi:https://doi.org/10.1016/j.scico.2020.102414}}.

\bibitem{scheer2020star3}
D.~Scheer, O.~Glodd, H.~G{\"u}nther, Y.~Duhr, A.~Schmid, Star3-eine neue
  generation der e/e-architektur, Sonderprojekte ATZ/MTZ 25 (2020) 72--79.
\newblock \href {https://doi.org/https://doi.org/10.1007/s41491-020-0056-5}
  {\path{doi:https://doi.org/10.1007/s41491-020-0056-5}}.

\bibitem{erickson1985qualitative}
F.~Erickson, et~al., Qualitative methods in research on teaching, Institute for
  Research on Teaching, 1985.

\bibitem{maxwell1992understanding}
J.~Maxwell, Understanding and validity in qualitative research, Harvard
  educational review 62~(3) (1992) 279--301.

\bibitem{bogner2019microservices}
J.~Bogner, J.~Fritzsch, S.~Wagner, A.~Zimmermann, Microservices in industry:
  insights into technologies, characteristics, and software quality, in: 2019
  IEEE international conference on software architecture companion (ICSA-C),
  IEEE, 2019, pp. 187--195.
\newblock \href {https://doi.org/https://doi.org/10.1109/ICSA-C.2019.00041}
  {\path{doi:https://doi.org/10.1109/ICSA-C.2019.00041}}.

\bibitem{fritzsch2019monolith}
J.~Fritzsch, J.~Bogner, A.~Zimmermann, S.~Wagner, From monolith to
  microservices: A classification of refactoring approaches, in: Software
  Engineering Aspects of Continuous Development and New Paradigms of Software
  Production and Deployment: First International Workshop, DEVOPS 2018, Chateau
  de Villebrumier, France, March 5-6, 2018, Revised Selected Papers 1,
  Springer, 2019, pp. 128--141.
\newblock \href {https://doi.org/https://doi.org/10.1007/978-3-319-67262-5_2}
  {\path{doi:https://doi.org/10.1007/978-3-319-67262-5_2}}.

\bibitem{tyszberowicz2018identifying}
S.~Tyszberowicz, R.~Heinrich, B.~Liu, Z.~Liu, Identifying microservices using
  functional decomposition, in: Dependable Software Engineering. Theories,
  Tools, and Applications: 4th International Symposium, SETTA 2018, Beijing,
  China, September 4-6, 2018, Proceedings 4, Springer, 2018, pp. 50--65.
\newblock \href {https://doi.org/https://doi.org/10.1007/978-3-319-99933-3_4}
  {\path{doi:https://doi.org/10.1007/978-3-319-99933-3_4}}.

\end{thebibliography}





\end{document}